\documentclass[]{aa}
\usepackage{amsmath}
\usepackage{txfonts} 
\usepackage{subfigure}
\usepackage{color}
\usepackage{bm}
\usepackage{natbib}
\usepackage{url}
\usepackage{tabularx}
\usepackage{graphicx}
\usepackage{txfonts}
\usepackage{longtable}
\usepackage{hhline}
\usepackage{arydshln}
\usepackage{multirow}
\usepackage{hyperref}

\newcommand{\hersc}{{\it Herschel}}

\newcommand{\msun}{$M_\odot$}

\newcommand{\mic}{$\mu$m}

\newlength{\pointwidth}
\begin{document}

\title{From cores to the inner envelope: Tracing variations in the dust emissivity index of the L1448 protostars}

  \author{A. Zuleta \inst{1},
            A. J. Maury \inst{1,2,3},
               L. Cacciapuoti \inst{4},
               J. M. Girart \inst{1, 5},
               V. Valdivia \inst{6,7},
               L. Testi \inst{8, 9},
               B. Ladjelate \inst{10}
}

\institute{
Institut de Ciències de l’Espai (ICE-CSIC), Campus UAB, Can Magrans S/N, E-08193 Cerdanyola del Vallès, Catalonia, Spain
\email{azuleta@ice.csic.es}
\and
ICREA, Pg. Lluís Companys 23, E08010 Barcelona, Spain
\and
Universit\'{e} Paris-Saclay, Université Paris Cit\'{e}, CEA, CNRS, AIM, 91191 Gif-sur-Yvette, France
\and
European Southern Observatory, Alonso de Cordova 3107, Vitacura, Region Metropolitana de Santiago, Chile
\and 
Institut d’Estudis Espacials de de Catalunya (IEEC), E-08034 Barcelona, Catalonia, Spain
\and 
Liant, 5 Av. Jean Jaurès, 34600, Bédarieux, France
\and 
Department of Physics, Nagoya University, Furo-cho, Chikusa-ku, Nagoya, Aichi 464-8602, Japan
\and 
Alma Mater Studiorum – Università di Bologna, Dipartimento di Fisica e Astronomia “Augusto Righi”, Via Gobetti 93/2, I-40129, Bologna, Italy
\and 
INAF-Osservatorio Astrofisico di Arcetri, Largo E. Fermi 5, I-50125 Firenze, Italy
\and 
Institut de Radioastronomie Millimétrique (IRAM), Avenida Divina Pastora 7, Local 20, E-18012 Granada, Spain
}

\abstract
{The (sub-)millimeter dust emissivity index ($\beta$) has been widely used as a proxy for dust grain evolution. In star forming regions, $\beta$ has been measured either on large scales using single-dish data or on small scales with interferometry, but these regimes have rarely been connected.}
{The goal of the study presented here is to assess, in a controlled environment, whether dust properties vary from the initial reservoir in the host cloud, all the way down to the disk-forming scales, using the L1448 Perseus clump and its protostars as template.}
{We determine the radial variations of the dust emissivity index at mm wavelengths, from cloud scales ($\sim 10^{4}~\rm au$) using single-dish IRAM-30m/NIKA2 dual-wavelength observations, to the inner envelopes ($\sim 10^{2}~\rm au$) using ALMA Band 3 and Band 6 observations. We analyze jointly both datasets to constrain the variations of $\beta$ within the clump.}
{This approach allows us to trace the dust emissivity index continuously from cloud scales down to the inner envelope, filling the gap that usually divides single-dish and interferometric measurements, revealing a continuous monotonic decrease of $\beta$ (from 2.0 to 0.3) over two orders of magnitude in spatial scales. We discuss potential physical processes and observational biases that could drive this observed correlation between the mm dust emissivity index and the scale probed in L1448.}
{Our work suggest that the dust properties evolve already at cloud scales ($30\,000>\rm{r}>15\,000~\rm au$), with a decrease of the dust emissivity index observed with gas density, and propagates further in the cores formed out of this material. In the L1448 protostellar envelopes probed at scales $\rm{r}<15\,000~\rm au$, we observe a relationship $\beta \propto n^{-0.17}$ at densities $n>10^5~\rm cm^{-3}$. If this change in $\beta$ is due to a dust grain size evolution, our results challenge current models describing dust growth, that cannot produce grains significantly larger than a dozen of microns at the typical protostellar densities probed in our observations.}

\keywords{Stars: protostars, formation, circumstellar matter -- ISM: dust -- Radio continuum: ISM}

\authorrunning{Zuleta, Maury, Cacciapuoti et al.}
\titlerunning{Tracing variations in the dust emissivity index of the L1448 protostars}
\maketitle
\nolinenumbers

\section{Introduction} 
Star formation happens preferentially within the filamentary structure of molecular clouds, where the cloud fragments into gravitationally bound cores. These cores can undergo collapse to form a protostar, eventually leading to a star and planetary system. During this evolutionary sequence the cloud material goes through rapid major transformations, gas and dust accrete from $0.1$~pc scales down to few tens to a few hundred au, with densities rising by several orders of magnitude. During this process, dust grains grow from the submicronic sizes typical of the interstellar medium \citep[ISM, $0.1~\mu$m,][]{MRN1977, Weingartner2001, Zubko2004} to centimeter sizes in protoplanetary disks \citep{Miyake1993, Testi2001, Ricci2010, Perez2012}. Interestingly, while dust evolution has been widely studied in circumstellar disks \citep[see e.g.,][for a recent review]{Birnstiel2024}, the properties of the dust grains that are incorporated into those disks and set their initial conditions are still relatively unknown. Because of the short lifetimes ($< 10^{5}$ yrs; \citealt{Evans2009, Maury2011, Kristensen2018}) and relatively low densities of the cores during the main accretion phase, especially when compared to the much longer time the dust sits in the protoplanetary disk (lifetimes of 1-10 Myr), most dust evolution models have ignored the early evolution of dust during the prestellar and protostellar stage. This blind corner is also due to the paucity and complexity of observations probing consistently dust properties at scales 0.1~pc in the dense ISM down to $\sim100~\rm au$ in the inner envelope.

Since at this stage both the disk forms and most of the stellar mass assembly takes place, embedded Class 0 protostars are the ideal environments for studying the pristine properties of dust grains during their travel from the dense cloud to the nascent disks \citep{Andre2000}. In particular, the dust grains play several fundamental roles during the main accretion phase, and their properties may affect the future evolution of both the star and the disk. For example, dust grains provide the surfaces on which complex organic molecules can form \citep{Garrod2006, Garrod2008}. In addition, electrons can be trapped on the surface of dust grains, affecting the ionization degree of the medium \citep{Marchand2016}. Grains also play an important role in regulating the collapse of the envelopes, by coupling the predominantly neutral gas to the local magnetic field, since grains become the primary charge carriers in these shielded environments, enabling efficient angular momentum transport from the infalling material \citep{Zhao2016, Zhao2018, Vallucci-Goy2024}. Moreover, dust particles are the building blocks of planets: grains grow from the micrometer-sized grains found in the ISM to pebbles, and eventually to the kilometer-sized bodies found in planetary systems \citep{Testi2014, Birnstiel2024}. The dust grains contained in the star-forming cloud, and in the protostellar envelopes, will set the initial conditions in the circumstellar disk, and may therefore influence the physical conditions for star and planet formation, making it crucial to characterize the dust evolution during these early stages of star formation.

Several independent observational diagnostics suggests the presence of evolved dust populations in dense clouds. For example, scattering of the interstellar radiation field by dust grains is sensitive to grain size. This phenomenon was observed in near-infrared (NIR) scattering, named as "cloudshine" \citep{Foster2006}, which traces the outer parts of molecular clouds. Mid-infrared (MIR) scattering, or “coreshine” \citep{Steinacker2010, Pagani2010, Steinacker2015}, has been observed in a few dense cores, suggesting a significant scattering also occurs in higher density conditions. Radiative transfer models that allowed grain growth, showed that it takes grains larger than typical $0.1~\mu$m ISM grains to scatter mid-infrared radiation \citep{Steinacker2010}, where grains up to $\sim 1~\mu$m in size were needed to explain the observed coreshine fluxes \citep{Andersen2013}. 

In the (sub-)millimeter regime, another probe of dust evolution is the dust emissivity (or dust opacity), $\kappa_{\nu}$, which is commonly parameterized as a power law, $\kappa_{\nu} \propto \nu^{\beta}$ \citep{Hildebrand1983}. In a population of grains with sizes spanning several orders of magnitude, an increase of large grains will lead to an increased emissivity at long wavelengths, and thus a decrease of the dust emissivity index, $\beta$. The dust emissivity index has often been derived from multi-frequency SED fitting of \hersc\ data and can be compared with independent estimates of the hydrogen column density, $N_{\mathrm{H}}$, obtained from near-infrared color excess measurements, $E(J-K_{s})$. Such comparisons have revealed a steeper than linear increase of opacity with column density relative to the calibration established for the high-latitude diffuse ISM, which is commonly interpreted as evidence for environmentally driven grain evolution. This behavior has been identified towards the Orion A molecular cloud \citep{Roy2013}, Perseus \citep{Okamoto2017}, and dark clouds in Ophiucus \citep{Li2024}. More recently, JWST ice-band spectroscopy toward dense clouds found that the observed ice-band absorption features are best reproduced when the dust size distribution extends to $\sim1\,\mu$m \citep{Dartois2024}.

Deeper into the molecular clouds and star-forming environments, evidence of dust growth have mainly come from measurements of the dust emissivity index, $\beta$, at (sub-)millimeter wavelengths. Several observational works have measured $\beta$ in the range of $1.0 \lesssim \beta \lesssim 2.5$ toward nearby star-forming clumps \citep{Sadavoy2013, Schnee2014, Chen2016, Pattle2025}. In the B213 filament, \cite{Bracco2017} reported radial variations of $\beta$ for Class 0/I protostars, while finding no radial trend for the prestellar core studied. Similarly, no radial variations of $\beta$ were found toward the prestellar core L1544 \citep{Chacon-Tanarro2017, Chacon-Tanarro2019}. Deeper into the scales probed by interferometric observations (from several thousands to few hundreds of au), a wide range of dust opacity spectral indices has been reported toward Class 0 protostars, with values spanning from $\beta \sim 0.5$ to $\beta \sim 2$ \citep[e.g.,][]{Kwon2009, Chiang2012, Galametz2019, Nozari2025, Cacciapuoti2025}. This diversity in $\beta$ may reflect intrinsic differences from source to source, as no clear indication of dependence with evolutionary stage was found \citep{Galametz2019, Cacciapuoti2025}. In addition, different dust grain models predict distinct dust opacity spectral indices arising only from variations in grain composition, structure, and stoichiometry \citep{Jones2017, Hensley2023}, which may partially explain the observed variations. Additionally, while state-of-the-art dust grain models can successfully reproduce the dust spectral index of the diffuse interstellar medium only including grains up to $0.7~\mu\mathrm{m}$ sizes \citep{Ysard2024}, they fail to reproduce values of $\beta < 1.0$ observed towards Class 0 objects \citep{Carpine2025}, where one way to reproduce such low $\beta$ would be if there were larger grains \citep{Testi2014, Birnstiel2024}. Taken together, these results indicate that the physical origin and spatial evolution of $\beta$ remain ambiguous, particularly due to the lack of observations consistently probing dust emissivity index across the full range of spatial scales involved in protostellar collapse.

In this work, we aim to investigate whether the properties of dust grains in protostellar environments are directly inherited from the diffuse ISM or whether dust evolution occurs during the assembly and subsequent collapse of cores into star-disk systems. Following the findings of low dust emissivity indices in the inner envelopes of protostars from the CALYPSO survey \citep{Galametz2019}, we obtained observations of the thermal dust emission with both ALMA and the IRAM-30m/NIKA2 camera. Our goal is to explore how the low $\beta$ values found with the CALYPSO survey in the inner envelopes are connected to the dust properties at larger scales in the surrounding envelopes and host filament. 

This paper is structured as follows. In Sec.~\ref{sec:Observations}, we provide details on the region and sources, the observations performed and data reduction process. In Sec.~\ref{sec:Results}, we present the methods used to derive the dust emissivity index and present our results. We discuss the implications of our findings in Sec.~\ref{sec:Discussion}. In Sec.~\ref{sec:Conclusions}, we summarize our findings and conclude.

\section{Observations}
\subsection{Sources}

We focus on L1448, which provides an ideal test case to study the dust emissivity index variations within a single environment. L1448 is a star-forming complex located at a distance of 293~pc in the Perseus molecular cloud, which is relatively isolated within the cloud. It has a total mass of $\sim$ 120\msun\ \citep{Sadavoy2014}. L1448 hosts several protostars at a similar evolutionary stage within a compact area, for which we have a unique dataset allowing us to trace $\beta$ from $\sim20\,000~$au down to $\sim400$~au, these sources are embedded in a common dense environment, thus they are expected to have a similar dust composition. The characteristic of the sources are summarized in Table~\ref{tab:source_properties}.

\begin{table}[ht!]
\caption{\label{tab:source_properties}Properties of the protostars in L1448 analyzed in this work.}
\centering
\resizebox{\columnwidth}{!}{
\begin{tabular}{lccccc}
\hline\hline
Name & $\alpha_{2000}$$^{(a)}$     & $\delta_{2000}$ & $T_{\mathrm{bol}}$$^{(b)}$ & $L_{\mathrm{int}}$ $^{(c)}$ & $M_{\mathrm{env}}$$^{(d)}$\\
     & [h:m:s]                     & [$^{\circ}$,',''] & (K)                        & ($L_{\odot}$)               & ($M_{\odot}$) \\
\hline
L1448-2A  & 03:25:22.40 & 30:45:13.26 & 43$^{(1)}$ & 4.7$^{(2)}$ & 1.8$^{(3)}$ \\
L1448-C   & 03:25:38.87 & 30:44:05.33 & 47$^{(1)}$ & 11$^{(2)}$  & 1.9$^{(1)}$ \\
L1448-NB1 & 03:25:36.38 & 30:45:14.77 & 57$^{(1)}$ & 3.9$^{(2)}$ & 4.9$^{(1)}$ \\
\hline
\end{tabular}
}
\tablefoot{$^{(a)}$ Equatorial coordinates measured from the peak positions of the 1.3 mm continuum emission maps derived in \citet{Maury2019}. $^{(b)}$ Bolometric temperature. $^{(c)}$ Internal luminosity. $^{(d)}$ Envelope mass.}
\tablebib{(1)~\citet{Sadavoy2014}; (2) \citet{Galametz2019}; (3) \citet{Enoch2009}.}
\end{table}

\label{sec:Observations}
\subsection{ALMA Observations}

We obtained ALMA data (project 2019.2.00167.S, PI: M. Galametz), using the Atacama Compact Array (ACA) with observations designed to probe dust emission from intermediate envelope scales for the three embedded protostars of L1448, and complemented them with archival ALMA high-resolution data allowing to characterize disk-like components in these systems. Our selection criteria for selecting archival data aimed at maximizing the common physical scales probed across different frequency bands, with comparable signal-to-noise ratio (SNR). This, in turn, consists in selecting data probing similar baselines in k$\lambda$. A summary of the observations is provided in Table \ref{tab:ALMA_observations}. In Appendix \ref{appendix:ALMA_data_reduction} we provide a description of the data reduction process for the individual datasets.

\subsection{NIKA2}
\label{sec:Iterative}

The New IRAM KID Array \citep[NIKA2;][]{Calvo2016, Adam2018} is a dual-band millimeter continuum camera of Kinetic Inductance Detectors (KID) installed on the IRAM 30m telescope and operating at 1.15 and 2.0 mm. NIKA2 has a field of view of 6.5\arcmin\ simultaneously at 1.15~mm and 2~mm, with a beam FWHM of $\sim12$\arcsec\ and $\sim18$\arcsec\, respectively. Thanks to its unprecedented mapping speed, large field-of-view and high angular resolution (for a single-dish millimeter instrument), NIKA2 is perfectly suited to map the dust thermal emission at relatively low brightness values with high SNR \citep[see][for a description of the instrument performance]{Perotto2020}.

\begin{figure*}
\centering
   \includegraphics[width=17cm]{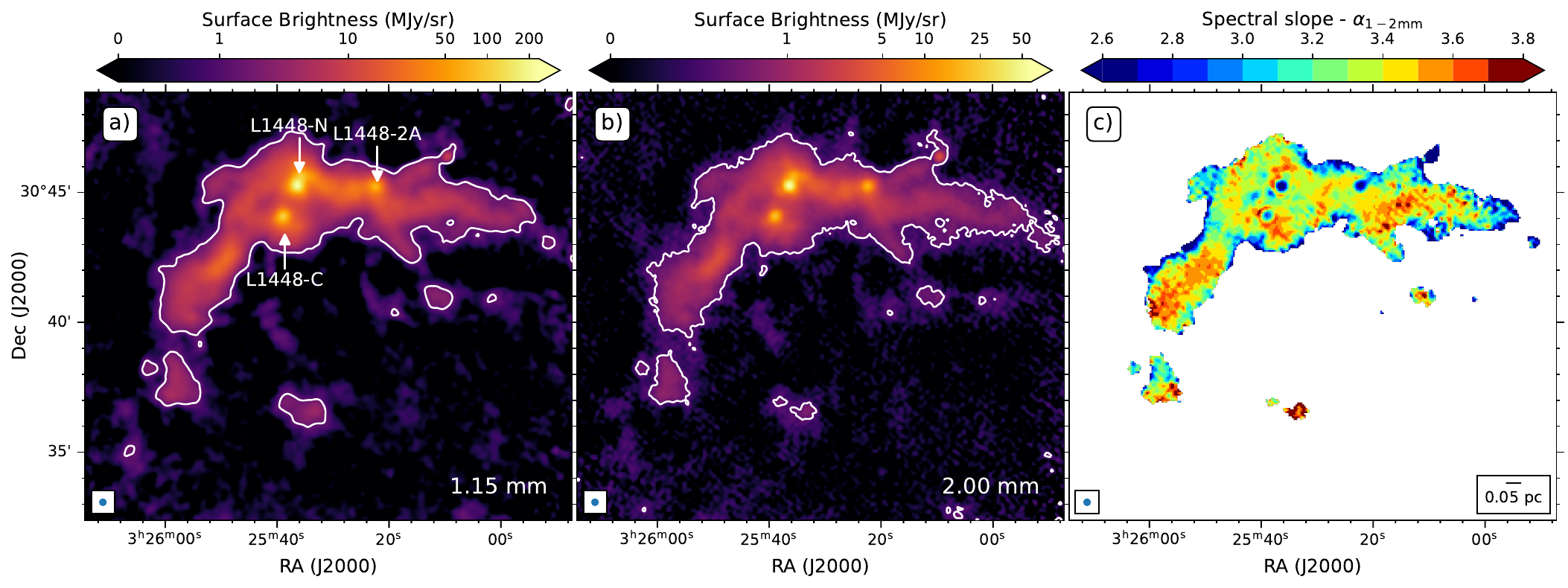}
     \caption{Maps of the thermal dust continuum emission obtained with NIKA2 towards L1448 at matched angular resolution. a) Dust continuum emission at 1.15~mm. b) Dust continuum emission at 2.00~mm. In panels a) and b), white contours indicate the 5$\sigma$ levels, where $\sigma$ corresponds to rms noise levels of 0.29 MJy~sr$^{-1}$ and 0.06 MJy~sr$^{-1}$ for the 1.15~mm and 2.00~mm maps, respectively. c) Spatial distribution of the observed spectral index, $\alpha_{\mathrm{1-2,mm}}$, across the L1448 complex. A drop in $\alpha_{\mathrm{1-2,mm}}$ is observed toward the three main protostellar cores. The beam size is shown in the lower left corner of each panel, and a physical scale bar of 0.05~pc is included in panel c).}
     \label{fig:nika_maps}
\end{figure*}

We used NIKA2 to map a 30\arcmin\ by 25\arcmin\ region to cover the whole L1448 clump above $A_{\rm V}$ of 8. We obtained observations of the region as part of project 123-18 (PI: B. Ladjelate \& M. Galametz) during February/March 2019. We performed a total of 30 individual on-the-fly maps towards the region, each with an approximated duration of 15 minutes, and a mapping speed of 70\arcsec/s. The source was at a mean elevation of $43^{\circ}$, and with a mean zenith opacity of 0.22 at 225GHz, which is equivalent to a precipitable water vapor of $\sim$3.7 mm. The reported flux calibration uncertainty was $5.5\%$ and $2.5\%$ for the 1.15 and 2 mm observations, respectively. We achieved an rms of 1.38 mJy/beam and 0.39 mJy/beam at 1.15 mm and 2 mm, respectively. We describe the calibration and map-making strategies in Appendix \ref{appendix:NIKA2}. The maps used to carry out the analysis are presented in Fig.~\ref{fig:nika_maps}.

\subsection{Ancillary Data}
\label{sec:Herschel}

In the FIR/submm regime, \hersc\ observations of L1448 were taken as part of the \hersc\ Gould Belt Survey \citep[HGBS;][]{Andre2010,Konyves2015}, with L1448 being part of the HGBS Perseus field. The PACS and SPIRE instruments aboard the \hersc\ Observatory have provided the maps of the thermal dust emission, sampling its spectral energy distribution (SED) from 70 to 500~\mic, with a spatial resolution ranging from $\sim1\,500$ au\ to $\sim10\,500$ au (at a 293 pc distance). An analysis of the Class 0 protostellar populations (identification, classification) of Perseus is presented in \citet{Sadavoy2014}. Core masses and temperatures have also been derived and analyzed in \citet{Pezzuto2021}. Published data products, column density and temperature maps from the HGBS were retrieved from the HGBS Archive \footnote{\url{http://www.herschel.fr/cea/gouldbelt/en/Phocea/Vie_des_labos/Ast/ast_visu.php?id_ast=66}}.

\section {Results}
\label{sec:Results}

\subsection{SED modeling}
\label{subsec:SED}

In the optically thin limit, the intensity of thermal dust emission is well described by a modified black body (MBB):

\begin{equation}
    I_{\nu} = I_{\nu}\left( \nu_{0} \right) \left( \frac{B_{\nu}(\nu, T_{\rm d})}{B_{\nu}(\nu_{0}, T_{\rm d})} \right) \left(\frac{\nu}{\nu_{0}}\right)^{\beta},
    \label{eq:MBB}
\end{equation}

where $\nu$ is the frequency, $I_{\nu}(\nu_{0})$ is the intensity at a reference frequency $\nu_{0}$, $B_{\nu}$ is the Planck function, $T_{\rm d}$ is the dust temperature, and $\beta$ is the dust emissivity index.

We constructed the SED using six continuum bands: \hersc\ PACS at 160 $\mu$m, SPIRE at 250, 350, and 500 $\mu$m, and NIKA2 at 1.15 and 2~mm. All maps were convolved to a common angular resolution of 36.3\arcsec, corresponding to the beam size of the \hersc\ 500 $\mu$m data, which has the lowest native resolution. The maps were then reprojected to a 14\arcsec-pixel grid. We excluded the 70 $\mu$m band because its emission may include contributions from very small grains that are not in thermal equilibrium, which has been reported toward Perseus \citep{Schnee2005, Schnee2008}, and thus cannot be reliably modeled with a MBB \citep{Martin2012}. We verified that including 70 $\mu$m band raises the fitted temperature by a median of $\sim2.9~\rm K$ and lowers $\beta_{\rm SED}$ by $\sim 0.3$, but since we use the SED fit only to constrain the temperature map and obtain $\beta_{\rm NIKA2}$ from the millimeter data, this has a negligible effect on our results (see Appendix~\ref{appendix:example_SED_fitting}).

Using the homogenized maps, we performed a pixel-by-pixel SED fitting to estimate the parameters of Eq. \ref{eq:MBB} via a $\chi^{2}$ minimization. We adopted flux uncertainties of $20\%$ and $10\%$ for PACS and SPIRE, respectively \citep{Konyves2015}, and $6\%$ and $3\%$ for NIKA2 $1.15~\rm mm$ and $2~\rm mm$ bands, respectively (see \citealt{Perotto2020}; see also Sec.~\ref{sec:Iterative}). Color correction factors were applied to all bands, with values of 1.01, 1.02, 1.01, and 1.03 for the \hersc\ 160, 250, 350, and 500~$\mu$m bands, and 1.005 and 0.997 for the NIKA2 1.15 and 2~mm bands, respectively \citep{Sadavoy2013, Perotto2020}. The parameter space was explored using a Markov Chain Monte Carlo (MCMC) approach to sample the posterior distributions of $I_{0}$, $T_{\rm d}$, and $\beta$. For each pixel, we adopted the median of the posterior distribution as the best-fit value, and the standard deviation as the associated uncertainty. We note that, to make sure we select reliable parameters, we masked out pixels with a signal-to-noise ratio below $5\sigma$ in both NIKA2 bands and retained only those pixels where the minimization performs well, producing a reduced $\chi^{2} \le 1$.

\begin{figure}
  \resizebox{\hsize}{!}{\includegraphics{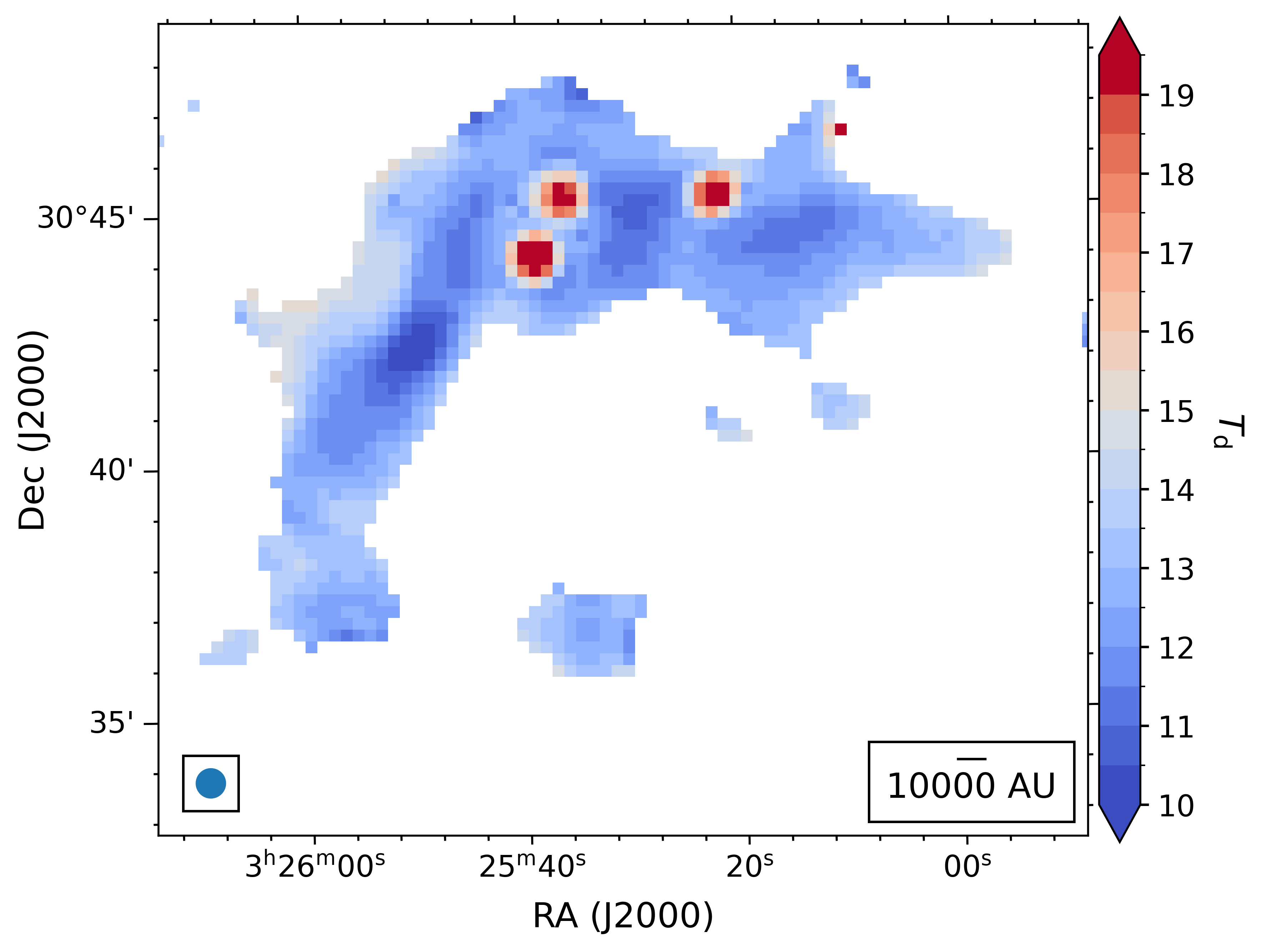}}
  \caption{Dust temperature map ($T_{\mathrm{d}}$) obtained from SED fitting. Black contours outline the region where the fit was performed (see Sec.~\ref{subsec:SED}). The synthesized beam is shown in the lower left corner, and a physical scale bar of $10\,000~\mathrm{au}$ (or equivalently $\sim 0.05~\mathrm{pc}$) is shown in the lower right.}
  \label{fig:Temperature_SED}
\end{figure}

The primary goal of the SED fitting is to obtain a robust estimate of the dust temperature $T_{\rm{d}}$, which is subsequently used in Sec.~\ref{subsec:NIKA2_emissivity} to correct the spectral index by the Planck function, and ultimately measure the dust emissivity index from the NIKA2 millimeter data. We stress that despite leaving $\beta$ as a free parameter, the derived $T_{\rm{d}}$ is close to the temperature maps published by the HGBS where they fix $\beta=2$ \citep[][see Appendix~\ref{appendix:temperature_comparison}]{Pezzuto2021}. The resulting dust temperature maps ($T_{\mathrm{d}}$), at a spatial resolution of $\sim10\,500$ au, is shown in Fig.~\ref{fig:Temperature_SED}. Within the fitted region, the dust temperature ranges from $10.6~\rm K$ to $23.1~\rm K$, with a mean value of $12.8\pm1.5~\rm K$. The temperature peaks coincide with the locations of the protostars, as expected from local heating, where the mean temperature reaches $17.2 \pm2.7~\rm K$. Conversely, the surrounding molecular cloud material is associated with colder regions.

\subsection{NIKA2 dust emissivity index map}
\label{subsec:NIKA2_emissivity}

\begin{figure*}
  \includegraphics[width=17cm]{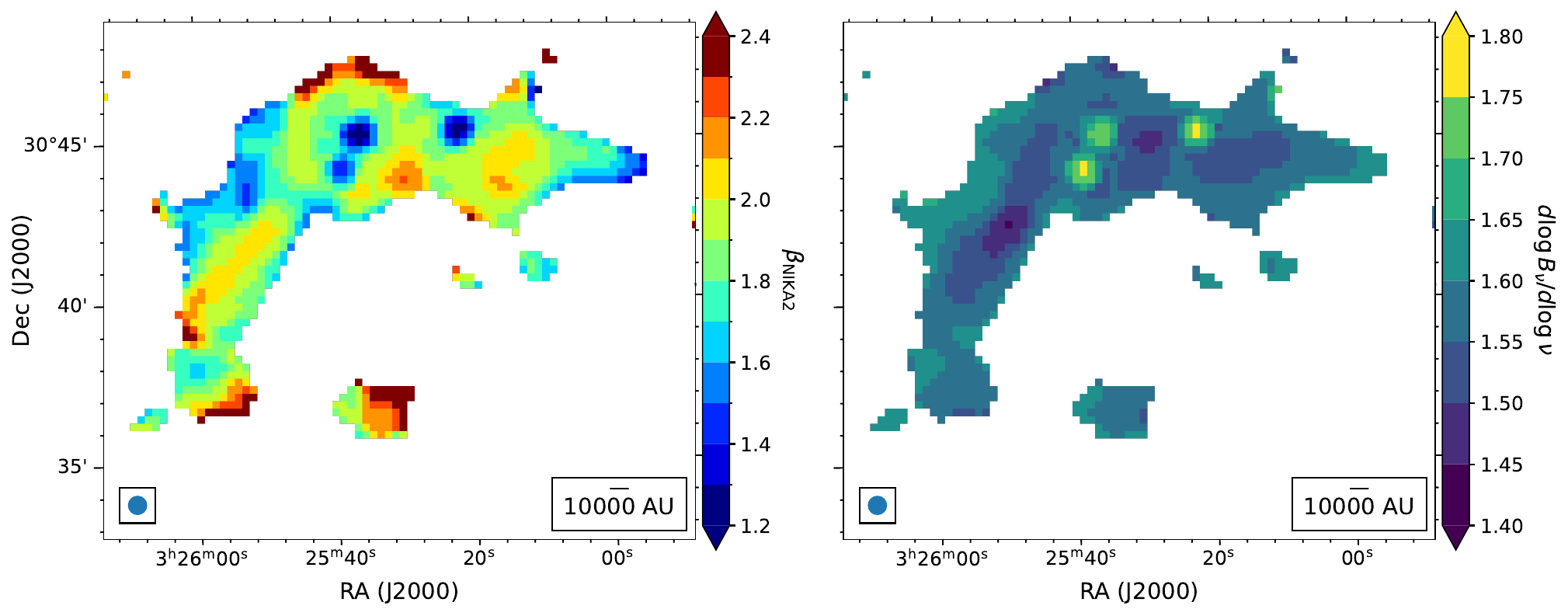}
    \caption{Left: Millimeter dust emissivity index map ($\beta_{\mathrm{NIKA2}}$) obtained by computing the spectral index using NIKA2 data and correcting with the dust temperature map obtained in Sec.~\ref{subsec:SED}. Right: Slope of the Planck function (second term on Eq.~\ref{eq:beta}) computed between $1.15~\rm mm$ and $2~\rm mm$ evaluated at $T_{d}$ (derived from SED fitting). On each plot, the synthesized beam is shown in the lower left corner, and a physical scale bar of $10\,000~\mathrm{au}$ (or equivalently $\sim 0.05~\mathrm{pc}$) is shown in the lower right.}
  \label{fig:beta_NIKA2}
\end{figure*}

The dust emissivity index at millimeter wavelengths was measured using the photometry in the dual-band NIKA2 observations at 1.15 and 2~mm. For dual-band instruments observing simultaneously, the emission in both bands can be assumed to be spatially coherent, minimizing errors due to relative calibration and filtering effects \citep{Rigby2018}. We first computed the millimeter spectral index, $\alpha$, as

\begin{equation}
    \alpha_{\nu_{1}-\nu_{2}} = \frac{\log(I(\nu_1) / I(\nu_2))}{\log(\nu_1 / \nu_2)},
\label{eq:alpha}
\end{equation}

where $\nu_{1}$ and $\nu_{2}$ correspond to the central frequencies of the NIKA2 bands, 260 and 150~GHz, respectively, and $I(\nu)$ is the measured intensity at each frequency. In the limit where $\nu_{1}$ and $\nu_{2}$ are sufficiently close, Eq.~\ref{eq:alpha} becomes the derivative $d\log{I_{\nu}}/d\log{\nu}$. The expression is valid as long as both frequencies are close. Assuming optically thin emission described by a modified blackbody (Eq.~\ref{eq:MBB}), the dust emissivity index can then be expressed as

\begin{equation}
    \beta = \alpha - \frac{d \log B_\nu (T_{\rm d})}{d \log \nu},
    \label{eq:beta}
\end{equation}

where the second term accounts for the deviation from the Rayleigh-Jeans approximation and depends explicitly on the dust temperature. The temperature information was taken from the SED-derived temperature map presented in Sec.~\ref{subsec:SED}. The corresponding uncertainties were derived consistently, by propagating errors in Eq.~\ref{eq:alpha} and Eq.~\ref{eq:beta} (see Appendix ~\ref{appendix:uncertainty_propagation}). We denote $\beta_{\rm NIKA2}$ as the dust emissivity index derived using NIKA2 observations. The resulting millimeter dust emissivity index map and the right part of Eq.~\ref{eq:beta} is shown in Fig.~\ref{fig:beta_NIKA2}. Across the L1448 star-forming clump, $\beta_{\mathrm{NIKA2}}$ ranges from 1.2 to 2.8, with a mean value of $1.98\pm0.2$. Note that the lowest $\beta_{\mathrm{NIKA2}}$ values coincide with the protostellar positions, where the dust temperature reaches its maximum.

\subsection{Interferometric visibility modeling}
\label{sec:visibility_mod}

The $\beta_{\mathrm{NIKA2}}$ map shown in Fig. \ref{fig:beta_NIKA2} shows a clear decrease of the millimeter dust emissivity index toward the protostellar positions, with values below 1.8. However, the 36.3\arcsec\ ($\sim10\,500~\rm au$) beam of the SED analysis (see Sec.~\ref{subsec:SED}) limits our capability to probe variations of $\beta$ at smaller scales. The ALMA data allows to explore further this observed decrease of $\beta$, probing the inner envelope scales NIKA2 cannot spatially resolve.

Because interferometers sample the complex visibility function at discrete locations in the $uv$-plane, a direct model fitting in the visibility space provides several advantages over image plane analysis. In the image domain, noise properties are not well characterized and extended sources may not be reliably recovered through image synthesis \citep{Pearson1999}. In contrast, each visibility has a well-defined variance, operating in the domain where observations where made. When the source structure can be accurately represented by a parametric model, visibility model fitting provides robust parameter estimates, and allows to disentangle the respective contributions coming from different components emitting at different spatial scales. Because the thermal dust emission from protostars is due to the combined contributions from the protostellar disk and the protostellar envelope, and these are typically difficult to disentangle in the image plane, we analyze the envelope dust thermal emission by modeling directly the interferometric visibilities. 

The brightness distribution of protostellar systems is well described by a combination of a compact disk component that can be approximated as a Gaussian and an extended power-law envelope \citep{Maury2019, Cacciapuoti2023, Cacciapuoti2025}:

\begin{equation}
    I_{\mathrm{tot}}(R) = I_{0}\cdot\mathrm{e}^{-\frac{R^{2}}{2\sigma^{2}}} + I_{\mathrm{env}}(R),
\end{equation}

where $R$ is the radial distance from the center of the image. The Gaussian component represents compact emission and is characterized by a peak intensity $I_{0}$, width $\sigma$, inclination $inc$, and position angle PA. The envelope is defined as a Plummer profile:

\begin{equation}
    I_{\mathrm{env}}(R) =
\begin{cases} 
I_{1}, & R < R_{\mathrm{in}} \\
\dfrac{I_{1}}{\left(1 + \left(\frac{R}{R_{\mathrm{in}}}\right)^{2}\right)^{\frac{p-1}{2}}}, & R_{\mathrm{in}} < R < R_{\mathrm{out}} \\
0, & R > R_{\mathrm{out}}
\end{cases}
\end{equation}

where $I_{1}$ is the envelope intensity at $R_{\mathrm{in}}$, $p$ is the power-law index, and $R_{\mathrm{in}}$ and $R_{\mathrm{out}}$ are the inner and outer radii. The envelope profile has a plateau for $R < R_{\mathrm{in}}$, a power-law behavior between $R_{\mathrm{in}}$ and $R_{\mathrm{out}}$, and it breaks at $R_{\mathrm{out}}$.

We modeled the ALMA visibilities with these parametric functions using \texttt{galario}\footnote{\url{https://mtazzari.github.io/galario/}}, which is a package that computes synthetic visibilities from the parametric model, sampled in the Fourier plane at the $uv$-point locations corresponding to the ones sampled by the interferometric observations and evaluates the $\chi^{2}$ of the model given the observed data \citep{Tazzari2018}. While the modeling is performed in the 2D $uv$-plane, for simplicity the 1D azimuthally-averaged radial visibility profiles for each source and frequency band are presented in Appendix~\ref{appendix:visibility_profiles}. We explored the parameter space using an affine-invariant Markov Chain Monte Carlo (MCMC) sampler \citep{Goodman2010} implemented in \texttt{emcee} \citep{Foreman-Mackey2013}.

The full model is described by ten parameters. For the Gaussian component, we fitted $I_{0}$, $\sigma$, $inc$, and PA. For the envelope, we fitted $I_{1}$, $R_{\mathrm{in}}$, $R_{\mathrm{out}}$, and $p$. Both components fixed to the same positional offset dRA and dDec relative to the phase center. All parameters were sampled linearly, except the intensities ($I_{0}$ and $I_{1}$), which followed log-uniform priors, with all priors centered on the envelope parameters reported by \citet{Maury2019} from their 1D modeling of CALYPSO visibilities. We employed 40 walkers and ran the chain for a total of $20\,000$ steps, discarding the first $12\,000$ as the burn-in phase. The best-fit model was defined as the median parameter values of their posterior distribution. The fit results for each source and Band are presented in Table \ref{tab:galario_results}.

\subsubsection{Compact component correction}
\label{subsubsec:compact_disk_correction}
Compact sources contribute to the integrated flux at all spatial scales probed by any baselines, as long as their size remains unresolved in front of the baseline length. While the compact disk component will contribute a much larger fraction of the flux at long baselines, its contribution at short baselines may remain non negligible, especially if the compact component (either disk or optically thick emission) has a different spectral index that will affect the one computed at shorter baselines. Corrections for the compact component were introduced  by \citealt{Galametz2019} and shown to be important when measuring $\beta$ from interferometric visibilities, and were used in \citealt{Cacciapuoti2023, Cacciapuoti2025}.

For L1448-C, we isolated the envelope emission by subtracting the best-fit Gaussian component (derived for each Band in Sec.~\ref{sec:visibility_mod}) from the visibility amplitudes, producing envelope-only visibility datasets. In the case of L1448-2A, the amplitude profiles do not show the characteristic disk saturation from the contribution of an important compact component. This suggests the contribution of the disk component to the millimeter flux, at these scales, must be very small, in agreement with the small disk size upper limit reported by \cite{Maury2019}. Yet, for consistency and robustness in removing any contribution from optically-thick emission, we subtracted from each band the 200k$\lambda$ flux, as unresolved sources contributes the same flux at all baselines \citep[see][]{Miotello2014}. We selected 10 k$\lambda$ as the shortest baseline and binned the compact component corrected data radially. 

In each $uv$-bin of the compact-corrected datasets, we derived the spectral index, $\alpha_{\rm ALMA}$, of the envelope thermal dust emission. For L1448-2A, where two frequencies are available, we directly computed Eq.~\ref{eq:alpha}. In the case of L1448-C, where three frequencies are available, we instead fitted a power-law to the visibility amplitudes as a function of frequency $I_{\nu}\propto\nu^{\alpha}$, and derived $\alpha$ from the best-fit slope. Uncertainties in $\alpha_{\rm ALMA}$ include both statistical and calibration contributions, with flux calibration dominating the uncertainty \citep[see][]{Cacciapuoti2023}. For L1448-N, we did not perform this analysis due to the intrinsic complexity of the source that would require to perform a 2D modeling of the multiple system contribution to the visibilities at different wavelengths \citep[see, e.g.,][]{Reynolds2021, Reynolds2024}, which falls beyond the scope of this paper. 

A potential contribution to the compact component may be due to free–free emission, which originates in ionized gas created by shocks near the base of the protostellar jet \citep[see][]{Anglada1998}. We estimated the free-free contribution at $3.0~\rm mm$ by extrapolating the VLA C-Band measurements of the VANDAM survey, adopting $F_\nu\propto\nu^{\alpha_{\rm ff}}$, with $\alpha_{\rm ff}$ equal to 0.52, and 0.99 for L1448-2A and L1448-C (\cite{Tychoniec2018b}). We then compared this value with the $3~\rm mm$ flux measured at $200~\rm k\lambda$ for both sources. For L1448-C the contribution is negligible ($5.6\%$). For L1448-2A the extrapolation shows that free-free could reach up to $19.4\%$ at $3~\rm mm$. In any case, this should not affect our dust spectral index, as free-free emission is expected to be compact, and thus removed by our subtraction of the flux at long baselines \citep{Tobin2013, Galametz2019}.

\subsubsection{Temperature correction}
\label{subsec:temperature_correction}
At large scales, the dust temperature can be directly obtained from the SED modeling. However, at small scales no such modeling is possible due to the lack of FIR photometry at comparable angular resolution, thus the dust temperatures at scales probed only by ALMA has to be modeled analytically. In the optically thin limit, the radial temperature profile of dust heated by a central protostar can be expressed as \citep[e.g.,][]{Terebey1993, Motte2001}:

\begin{equation}
    \label{eq:temp_profile}
    T(r) = 38 \left ( \frac{L_{\mathrm{int}}}{L_{\odot}} \right)^{0.2}\left( \frac{r}{100~\mathrm{au}} \right)^{-0.4},
\end{equation}

where $L_{\mathrm{int}}$ is the internal luminosity in $L_{\odot}$. We adopt values of $L_{\mathrm{int}} = 4.7~L_{\odot}$ for L1448-2A and $11~L_{\odot}$ for L1448-C \citep{Galametz2019}. Using these temperature profiles, we converted the spectral index measured for ALMA, $\alpha_{\mathrm{ALMA}}$, into the dust emissivity index $\beta_{\mathrm{ALMA}}$ using Eq.~\ref{eq:beta}.

\subsection{Dust emissivity index profiles}
\label{subsec:radial_dust_emissivity_profiles}

To construct a multi-scale view of the dust emissivity index ($\beta$), we derived a NIKA2-based emissivity index, $\beta_{\mathrm{NIKA2}}$ (see Sec.~\ref{subsec:NIKA2_emissivity}), then compared them with the large-scale maps with small-scale measurements obtained through ALMA visibility modeling (see Sec.~\ref{sec:visibility_mod}). 

At large scales, we extracted radial profiles from the dust spectral index $\beta_{\mathrm{NIKA2}}$ by centering on the 1.15 mm peak emission position of each source and computing azimuthally averaged values within concentric circular annuli. For the ALMA data, $\beta_{\rm ALMA}$ values measured as a function of $uv$-distance were converted to physical scales\footnote{Transformation from uv-distances to physical scale was done by computing $\mathcal{D}$ = 1.029 / ($uv$-distance) $\cdot$ 206265 $\cdot\ d$, where 1.029 corresponds to the constant factor preceding the FWHM expression of the Airy disk, 206265 is the factor to transform from radians to arcsecond, and $d$ the distance in parsec.}, to allow for comparison with the single-dish profiles.

\begin{figure*}
\sidecaption
  \includegraphics[width=12cm]{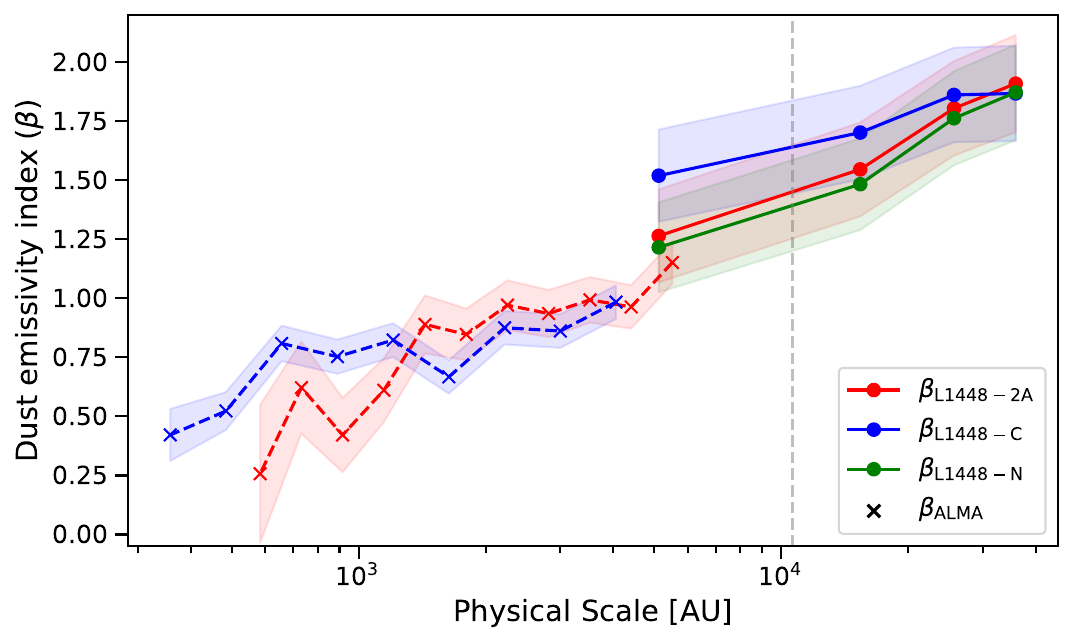}
     \caption{Multi-scale dust emissivity index profiles extracted from ALMA envelope visibilities (small to intermediate scales, using 1.3 and 3.3 mm data) and from NIKA2 maps (large scales, using 1.15~mm and 2~mm data). Symbols show the mean $\beta$ in each annulus, the shaded regions indicate the standard deviation of the pixel values within the annuli. Values measured at ALMA scales are indicated by "x" markers connected by dashed lines, while the large-scale profiles are shown as dots joined by solid lines. The profiles extend out to the radius at which $\beta$ reaches a plateau. The gray dashed line represents the spatial extent of the convolved 36.3\arcsec~ Herschel beam.}
     \label{fig:beta_profile}
\end{figure*}

The resulting radial $\beta$ profiles for each protostellar source are shown in Fig.~\ref{fig:beta_profile}. Over almost two orders of magnitude, we observe a monotonic evolution of the dust emissivity index, $\beta$, with decreasing spatial scale from the relatively homogeneous values at large scales ($\beta \sim 1.9$) to much smaller values measured inside the inner envelope ($\beta \sim0.3$ at scales of $\sim400$ au). This analysis allows to establish the spatial evolution of the mm dust emissivity index of dust particles, continuously and at all scales from the parent filament (0.2 pc) down to the inner envelopes of the protostars forming from this material, to a few hundreds of au scales. In the following, we discuss the findings and propose possible interpretation from them.

\section{Discussion}
\label{sec:Discussion}
With the aim of characterizing the dust emissivity index across multiple spatial scales, we presented in the previous section an approach combining different instruments and techniques: on large spatial scales, $\beta$ is derived from the NIKA2 millimeter spectral index, corrected by the dust temperature map obtained from SED fitting. On the smaller spatial scales probed with ALMA, $\beta$ is inferred from the envelope-only visibilities. In the following discussion, we focus on the factors that can account for the observed variations in the dust emissivity index, emphasizing how radial changes may reflect dust evolution and/or variations in the local physical conditions.

\subsection{Dust emissivity index: From cloud to inner envelopes}
\label{ssec:dbeta}
Our observations with NIKA2 show that, at large scales (0.2~pc), the dust emissivity index from the parent material in the L1448 filament is rather homogeneous in the surrounding of all three protostars, with typical $\beta_{\mathrm{NIKA2}} \sim 1.9$. We stress that this value is in the upper end of the $\beta$ distribution reported for the diffuse ISM \citep{Planck2014, Ysard2024}. A notable feature from our L1448 observation is the monotonic decrease of $\beta$ at all scales probed, as one gets from the filament into the inner cores: we find that the dust emissivity index $\beta$ decreases from 1.9 to $\sim0.5$ when spanning smaller and smaller scales over two orders of magnitude (down to $\sim400~$au scales probed with ALMA). While localized decrements at the positions of the three Class~0 protostars were already seen in the NIKA2 spectral index maps (see Fig.~\ref{fig:nika_maps}c), our analysis shows that this sharp decrease in spectral indices is not an effect of the increment of dust temperature alone, as the dust emissivity index map derived from NIKA2 maps ($\beta_{\mathrm{NIKA2}}$, see Fig.~\ref{fig:beta_NIKA2}) also reveals these drops. Localized $\beta$ variations toward these sources were also reported by \citet{Chen2016}, who combined \hersc\ and JCMT observations to derive $\beta$ and $T_{\mathrm{d}}$ maps through SED fitting. Their temperature and $\beta$ values are consistent with our measurements, although the decrease in $\beta$ towards the protostars is less pronounced. This difference likely arises from the inclusion of the NIKA2 mm wavelengths in our SED fitting procedure, which tends to increase the spatial gradient of $\beta$ towards the protostars.

\cite{Bracco2017} studied the radial slope in $\beta$ ($d\beta/dr$) for three sources in the B213 filament in Taurus using single-dish NIKA data, and observed a different behavior for $d\beta/dr$ for a prestellar core and protostellar sources. For the prestellar core, they found a radial slope ($d\beta/dr$) consistent with zero and having a mean value of $\bar{\beta} = 2.4\pm0.03$. For the protostellar sources K04166 and K04169, they measured positive gradients of $d\beta/dr = 0.136 \pm 0.06~\mathrm{(1000~au)}^{-1}$ and $0.129 \pm 0.06~\mathrm{(1000~au)}^{-1}$, respectively. Following \cite{Bracco2017}, we report the best fitting values obtained by performing a linear fit to the $\beta$ profiles obtained from single dish observations in Table. \ref{tab:radial_beta_linear_fit}. Our slopes range from $0.012\pm0.003~\mathrm{(1000~au)}^{-1}$ to $ 0.022\pm0.003~\mathrm{(1000~au)}^{-1}$, and are less steep than those reported by \cite{Bracco2017} for the protostellar sources, possibly reflecting differences in their dust evolution. Moreover, \cite{Rigby2018} studied the dust emissivity index using NIKA toward two Infrared Dark Clouds (IRDCs) in the galactic plane (at $\sim1$ pc scales), SDC18 which is located at the edge of the W39 H II region, and SDC24 with no nearby H II region. IRDCs are valuable targets because their low level of embedded star formation implies that the initial conditions of the material are still largely intact. Similarly as in the case of \citet{Bracco2017} for the prestellar core, \cite{Rigby2018} found no significant radial variations of $\beta$ in the IRDCs, but reported different means in the two distinct clouds. Toward the inner envelopes of young protostars, evidence of spatial variations of $\beta$ were reported by \cite{Galametz2019}, where they showed a radial decrement for 9 of 12 sources in the CALYPSO survey, with the values of $\beta$ measured ranging from 0.44 to 1.41 at 500~au. Building on the CALYPSO results, \citet{Cacciapuoti2025} applied a similar analysis to 11 sources of the ALMA FAUST Large Program. They measured a variety of $\beta$ values ranging from 1 to 1.8, where they found $d\beta/dr>0$ only for a subset of sources, others with flat profiles consistent with $d\beta/dr\sim0$. Note that while the L1448-C and L1448-2A protostars were analyzed in the CALYPSO survey \citep{Galametz2019}, our observations provide higher sensitivity and present the first case in which such radial variations are measured consistently across both large and small spatial scales for the same sources.

\begin{table}[ht]
\caption{Parameters of the linear fit, $\beta_{\mathrm{NIKA2}} \left (  r \right ) = m \times r + \beta_{\text{in,\ NIKA2}}$.}
    \centering
    \begin{tabular}{c c c} 
    \hline\hline
         System & Slope ($m$)                    & Inner $\beta$ value ($\beta_{\mathrm{in,\ NIKA2}}$) \\
                &             [(1000 au)$^{-1}$] &  \\
    \hline
         L1448-2A & 0.021 $\pm$ 0.003 & 1.19 $\pm$ 0.07 \\
         L1448-C  & 0.012 $\pm$ 0.003 & 1.49 $\pm$ 0.07 \\
         L1448-N  & 0.022 $\pm$ 0.003 & 1.13 $\pm$ 0.06 \\
    \hline
\end{tabular}
\tablefoot{Fit is performed to the large-scale $\beta$ profiles shown in Fig.~\ref{fig:beta_profile}. Here, $\beta_{\text{in,\ NIKA2}}$ represents the dust emissivity index at the inner boundary of the radial range considered for the fit.}
\label{tab:radial_beta_linear_fit}
\end{table}

Low dust emissivity indices have also been reported at a range of spatial scales for protostellar cores in the OMC 2/3 region. \citet{Mason2020} measured low $\beta$ on $\sim0.1~\rm pc$ scales using single-dish data. In their interferometric study of 6 Orion cores, \citet{Nozari2025} find cores with $\beta$ between $-0.16$ and $1.45$, notably lower than their measurements from single-dish data \citep{Mason2020}. This is consistent with our results in the sense that the Orion cores also exhibit lower $\beta$ than their parent cloud when observed at smaller scales. Our analysis of the interferometric data differs in two aspects. First, we subtract a compact component from the visibilities before measuring the spectral index, since emission from an unresolved, optically thick (or partially optically thick) component would lower the measured $\alpha$ and hence $\beta$ at all baselines (see Sec.~\ref{subsubsec:compact_disk_correction}). \citet{Nozari2025} report a substantial compact disk contribution in their sources, they do not remove this contribution in their analysis.  Second, in the absence of temperature information they adopted $\beta=\alpha-2$, which is the Rayleigh--Jeans limit of Eq.~\ref{eq:beta}, for example at a temperature of $20~\rm K$, this approximation leads to underestimating $\beta$ by $\sim0.25$ \citep[see][]{Carpine2026}.

While previous studies have measured $\beta$ at either large or small scales, our work provides the first radial measurements of the dust emissivity index across order of magnitudes in scales, from cloud scales down to the inner envelope for sources within the same environment and at similar stages of their evolution. Here, $\beta$ is measured radially across spatial scales using both single-dish and interferometric data, this shows that two young protostars that form from a common environment, with a unique dust emissivity, both exhibit a strong and monotonic decrease of their dust emissivity index as going deeper into their envelopes, probing a change of $\beta$ from $\sim2.0$ at $\sim20\,000$ au down to $\sim 0.5$ at $\sim400$ au.

\subsection{Physical origin of $\beta$ variations}
\label{ssec:origbeta}
In this section we will explore what are the physical mechanisms that can be responsible for the observed spatial variations of the dust emissivity index, $\beta$.

\subsubsection{Variations due to dust composition}
Variations in the composition of dust grains may also contribute to the observed variations in the dust emissivity index $\beta$ measured at millimeter wavelength. However, laboratory measurements of interstellar dust analogues have shown that reproducing very low millimeter emissivity indices, with $\beta_{\mathrm{mm}} < 1$, remains challenging \citep[see e.g.,][]{Demyk2017a,Demyk2017b}. For dense cores forming within the same molecular cloud, we expect rather an uniform dust composition given the common material reservoir. This is reflected in the $\beta$ measured at large scale for our sources (see Fig.~\ref{fig:beta_profile}). 

Even if dust grain of different composition were to become spatially segregated, we argue against this possibility based on timescale arguments. If we compute the crossing time for a dust grain inside the cloud: $t_{\mathrm{cross}}\sim L/v$, where $L$ is a characteristic size and $\rm{v}$ is the velocity of the grain, we assume $v$ equal to the non-thermal velocity dispersion. For L1448, \citet{Pineda2026} measured a non-thermal velocity dispersion of $0.18~\rm km/s$ from NH$_{3}$. Assuming $L=10\,000$~au as the typical size of a protostellar core, this gives $t_{\mathrm{cross}}\sim2.6\times10^{5}$~yr. This crossing time is on the same order of the Class 0 lifetime ($\sim10^{5}$~yr), indicating that dust grains have sufficient time to move significantly within the system. Therefore, protostars forming from the same cloud should not show significant differences in their dust composition that could alter the dust emissivity index.

\subsubsection{Opacity effects}
In optically thick regions, the emergent intensity approaches the blackbody limit ($I_{\nu} \propto B_{\nu}(T)$) and the spectral slope $\alpha$ no longer traces the variations of the dust opacity with wavelength. As a consequence, attempting to derive a dust emissivity index from optically thick emission results in artificial low values of $\beta$. However, at the scales probed by the single-dish observations ($\sim10\,000~$au, or 0.05 pc for typical column densities of $N_{\rm H_{2}}\sim10^{22}~\rm cm^{-2}$), we expect dust emission to be optically thin at submillimeter wavelengths. In the Perseus molecular cloud, optically thick emission has been reported primarily at $70~\mu$m \citep{Schnee2005,Schnee2008}, while longer wavelengths remain well within the optically thin regime. In particular, measurements toward L1448 indicate optical depths of $\tau_{300~\mu\mathrm{m}} \lesssim 3 \times 10^{-2}$ \citep{Chen2016}, implying even lower optical depths at millimeter wavelengths. At smaller spatial scales, we directly evaluate the optical depth of the envelope-only visibilities as function of baseline by combining the measured flux density with the corresponding dust temperature. We estimate the optical depth at 1.3~mm, $\tau_{1.3~\mathrm{mm}}$, using:

\begin{equation}
    \frac{S_{1.3~\mathrm{mm}}}{\Omega} = \left(1 - e^{-\tau_{1.3~\mathrm{mm}}}\right) B_{1.3~\mathrm{mm}}(T_{\mathrm{d}}),
\end{equation}

where $S_{1.3~\mathrm{mm}}$ is the flux density at 1.3~mm, $\Omega$ is the solid angle associated with the physical scale probed by each baseline, and $B_{1.3~\mathrm{mm}}(T_{\mathrm{d}})$ is the Planck function evaluated at the dust temperature. Using this approach, we find maximum optical depths of $\tau_{1.3~\mathrm{mm}} \simeq 0.002$ and $0.004$ for L1448-2A and L1448-C, respectively, across all baselines. These values confirm that the emission is in the optically thin regime at all spatial scales probed by our observations (above $\gtrsim 400~$au). Similarly, \cite{Galametz2019} evaluated the optical depth of the envelope-only emission of the CALYPSO sample and found that most of the sources are optically thin at millimeter wavelengths, except for the very central emission ($<200~$au) of three sources (IRAS~4A, IRAS~4B, and SerpM-SMM4), all of which have more massive envelopes than in our case ($M_{\mathrm{env}} = 12.2$, $4.7$, and $5.4~M_\odot$, respectively). Since our analysis does not target envelopes of comparable mass and we correct for an optically thick component, optically thick emission cannot account for the observed variations in $\beta$.

\subsubsection{Ice mantles}
In molecular clouds, the freeze-out of gas species is effective, creating ice layers on top of the dust grains. These ice layers remain on the grain surfaces until the temperature becomes sufficiently high for desorption back into the gas phase. The optical properties of dust are affected by the presence of ice mantles, which can modify the measured dust emissivity index. The net effect on the total dust emissivity depends on the relative emissivities of the bare grain and ice mantle. If the grain has a higher emissivity than the ice, the resulting emissivity will be lowered \citep[see e.g.,][]{Aannestad1975, Pollack1994, Ossenkopf1994, Carpine2025}. Different molecular species sublimate at different temperatures, with CO serving as a benchmark at these scales, because it has a relatively high abundance and it is the ice with the lowest desorption temperature \citep[$T \sim30~\mathrm{K}$,][]{Jorgensen2013, Jorgensen2015}. On large scales, our SED fitting process yields a maximum dust temperature of $23.1~\mathrm{K}$. Using Eq.~\ref{eq:temp_profile}, the maximum temperature expected at the smallest scales we probe is $\sim28~\mathrm{K}$ and $\sim35~\mathrm{K}$ for L1448-2A and L1448-C, respectively. Therefore, across all scales we probe here, we expect the dust temperature to be below CO sublimation temperature at most scales. One caveat is that Eq.~\ref{eq:temp_profile} may not fully reflect the true temperature structure of the envelope. We discuss this effect further in Sec.~\ref{ssec:caveats}. Indeed, chemical modeling of L1448-C indicates a freeze-out radius for CO at $r_{\mathrm{snow}} = 730~$au (equivalent to 1500~au scales) \citep{Anderl2016}. This potentially only affects the innermost part of our dust emissivity index profile, and is unlikely to explain the global change in dust emissivity index we observe across scales in our profiles.

\subsubsection{Dust size evolution}
\label{subsubsec:Dust_size_evolution}
It is widely admitted, and often used as a dust size probe, that enhanced long‑wavelength emissivity can be interpreted as coagulation into fluffy aggregates that increase $\kappa_{\nu}$ at mm wavelengths, as dust emissivity is sensitive to the maximum grain sizes, with a systematic decrease in $\beta$ expected when the grain population extends to particles with sizes $\gtrsim 100~\mu$m \citep{Natta2004, Draine2006, Testi2014, Ysard2019}. Therefore, the low dust emissivity indices measured in our data could easily be interpreted as an evolution of the grain size distribution with gas density. Yet, such a widely-spread interpretation produces some tension with the expectations from dust evolution models, that we discuss here below.

Early models of dust coagulation in molecular clouds have shown that the degree of grain growth depends primarily on the cloud lifetime and density \citep{Ormel2009}. However, even the choice of the dust evolution model can predict significant differences for grain sizes under the same physical conditions. For example, for $t=10^{5}$~yrs and $n=10^{7}~\rm cm^{-3}$, the predictions can range from 2~$\mu$m \citep{Silsbee2022} to as large as 0.4~mm \citep{Ormel2009}. This growth can be further enhanced for longer timescales. Observationally, prestellar core lifetimes have been estimated to span $2-5$ free-fall timescales based on young stellar objects lifetimes \citep{Ward-Thompson2007}, while recent simulations suggest a $1-10$ free-fall timescales range \citep{Offner2025}. Interestingly, despite these relatively long lifetimes, no clear radial decrease in $\beta$ has been observed toward prestellar cores \citep[see, e.g.,][]{Chacon-Tanarro2017, Bracco2017}, implying that evolution in the dust population in the starless phase is not significant to modify $\beta$ but seems to be triggered once the systems enters the Class 0 stage.

To assess whether the densities in our sources are sufficient to enable significant growth of the initial population of dust grains, we estimated the number density for L1448-C and L1448-2A down to ALMA scales. For this, we combined the large-scale $N_{\mathrm{H}_{2}}$ maps described in Sec.~\ref{sec:Herschel}, which define the density normalization, with the envelope parameters derived from the visibilities, which constrain the profile shape at small scales. 

First, we derived the number density at a reference radius outside the region probed by ALMA. We considered the first point of the radial $N_{\mathrm{H}_{2}}$ profile outside the \hersc\ 36.3\arcsec\ beam, at $R_{\rm out}\sim15000~\rm au$. At this radius $N_{\mathrm{H}_{2}} = 2.99\times10^{22}~\rm cm^{-2}$ and $N_{\mathrm{H}_{2}} = 2.26\times10^{22}~\rm cm^{-2}$ for L1448-C and L1448-2A, respectively. Assuming a line-of-sight depth $l$ equal to 3\arcmin\ (roughly the filament width), we obtain $n_{0} \simeq N_{\mathrm{H}_{2}}/l$.

Second, we assumed that the density of the envelope follows a Plummer-like profile, 
\begin{equation}
n(r)=\frac{n_{\rm c}}{\left[1+(r/R_{\rm i})^2\right]^{p/2}},
\end{equation}

we use the values of the envelope fits reported in \citet{Maury2019} (model PG231), chosen for consistency in the methods used to derive the envelope parameters. Since these fits only constrain the combined $p+q$ of the density and temperature profiles, we obtained $p$ by adopting $q=0.4$, as in Sec.~\ref{subsec:temperature_correction}. We then fixed the normalization $n_{\rm c}$ by requiring $n(R_{\rm out})=n_{0}$. Finally, we used the resulting profiles to obtain a density at each scale probed by ALMA, allowing us to relate $\beta_{\rm ALMA}$ directly to $n_{\mathrm{H}_{2}}$. Given the assumptions involved, these densities should be regarded as order of magnitude estimates.

\begin{figure}
  \resizebox{\hsize}{!}{\includegraphics{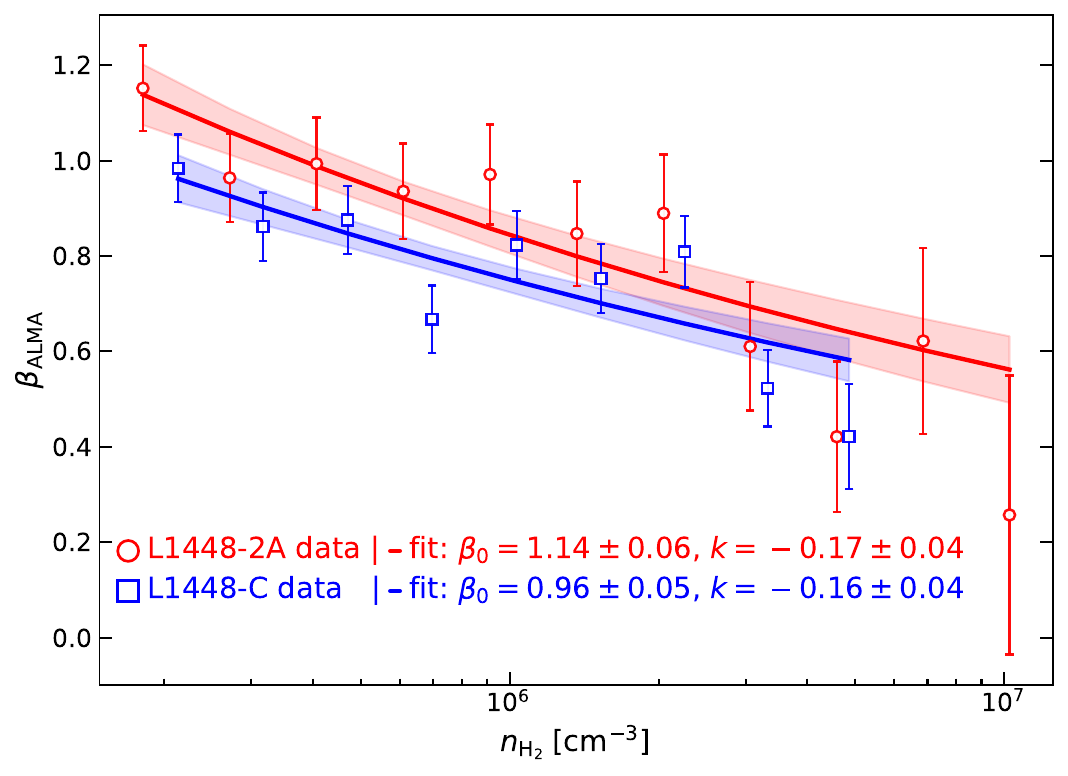}}
  \caption{Dust emissivity index derived at ALMA scales, $\beta_{\rm ALMA}$, as a function of number density, $n_{\mathrm{H_{2}}}$, reconstructed from the envelope parameters. Data points are shown as red circles and blue squares for L1448-2A and L1448-C, respectively). A normalized power law of the form $\beta(n_{\mathrm{H_2}}) = \beta_0 \cdot (n_{\mathrm{H}_{2}}/n_{\mathrm{min}})^{k}$ is fitted to each source, with the best-fit results shown as solid lines. The parameter $n_{\mathrm{min}}$ is a normalization factor. The best-fit parameters are indicated in the lower left corner.}
  \label{fig:beta_vs_nH2}
\end{figure}

We show this estimation in Fig.~\ref{fig:beta_vs_nH2}. At large scales (high $\beta_{\rm ALMA}$), we find $n_{\mathrm{H}_{2}}\sim1.8\times10^{5}~\mathrm{cm^{-3}}$ at $\sim 5\,500$ au for L1448-2A and $n_{\mathrm{H}_{2}}\sim2.1\times10^{5}~\mathrm{cm^{-3}}$ at $\sim 4\,000$ au for L1448-C. Toward smaller scales (lower $\beta_{\rm ALMA}$), the density increases to $n_{\mathrm{H}_{2}}\sim1.0\times10^{7}~\mathrm{cm^{-3}}$ at $\sim 600$ au for L1448-2A and $n_{\mathrm{H}_{2}}\sim4.9\times10^{6}~\mathrm{cm^{-3}}$ at $\sim 400$ au for L1448-C. Over this nearly two orders of magnitude increase in number density, both sources exhibit a similar behavior of $\beta$, with a tentative drop in $\beta$ once $n_{\mathrm{H_{2}}}\sim2\times10^{6}~\mathrm{cm^{-3}}$. Fitting a power law to $\beta$ as a function of number density results in nearly identical exponents for the two sources.

This similarity is remarkable, because although L1448-2A and L1448-C originate from the same initial reservoir and share similar large-scale environments, their inner properties such as luminosity and bolometric temperature differ, yet the evolution of the dust emissivity index $\beta$ with density is nearly identical in both protostars. The similarity of the profiles in these two systems, along with the fact that less evolved systems do not exhibit a radial evolution of $\beta$, seem to indicate that the physical process responsible for the change in the dust emissivity index must be fast and happen at the beginning of the protostellar phase, reaching a steady-state, thus no strong dependency of the spatial evolution of $\beta$ is found with stellar properties.

In terms of the models, only within the very inner regions ($r<700$ au) of the L14448 protostars reach the typical values at which dust coagulation models predict the formation of aggregates of $\sim100~\mu\mathrm{m}$ \citep{Ormel2009}.  Moreover, $\beta$ values as low as $\sim 0.3$ would require a large amount of grains with sizes $>100~\mu$m as suggested by many state-of-the-art dust models \citep{Testi2014, Ysard2019}. More recent work incorporating the treatment for the dust charge further support this scenario. Magneto-hydrodynamic (MHD) simulations by \citet{Valluci-Goy2025} show that dust density enhancements of an order of magnitude can develop within a few Alfvén crossing times for $1~\mu$m grains, and up to factors of $\sim30$ for $10~\mu$m grains. Because higher local densities reduce the coagulation timescales, such clumping accelerates grain growth, enabling the rapid production of large aggregates at envelope densities early in the protostellar phase. These models, although potentially providing a plausible physical mechanism to reach large grains sizes at turbulence dissipation scales, still need to be explored in their efficiency to grow a fair fraction of the dust mass at envelope scales.

\subsection{The $\beta$ and $T$ relation}
\label{subsec:beta-T}
Laboratory measurements have also shown that the dust emissivity changes also depends on temperature \citep{Draine1984, Coupeaud2011, Demyk2022}. To quantify the effect of temperature on the inferred emissivity index $\beta$, we use the optical constants from \citet{Demyk2022}. Here, they provide measurements on submicronic grains for several amorphous silicate analogues, including extrapolations beyond the probed experimental range both in temperature and wavelength. We adopt their X35 sample, which has a composition close to Mg$_{2}$SiO$_{4}$. We computed the absorption coefficient $\kappa_\nu$ at 1 and 2 mm for temperatures between 10 and 20 K and between 10 and 50 K, for which we derive the corresponding $\beta(T)$ values. Over the 10 to 20 K temperature range, $\beta$ only varies between 1.8 to 1.7 with temperature. Over the 10 to 50 K range it decreases only from approximately 1.8 to 1.5. This variation contrasts with the significantly large range of $\beta$ values measured (from $2.0$ down to $\beta \sim 0.3$ over a $\sim10~\rm K$ to $\sim35~\rm K$ temperature range), suggesting that changes in the optical properties due to temperature alone cannot account for the observed decrement of $\beta$ toward smaller scales.

\begin{figure}
  \resizebox{\hsize}{!}{\includegraphics{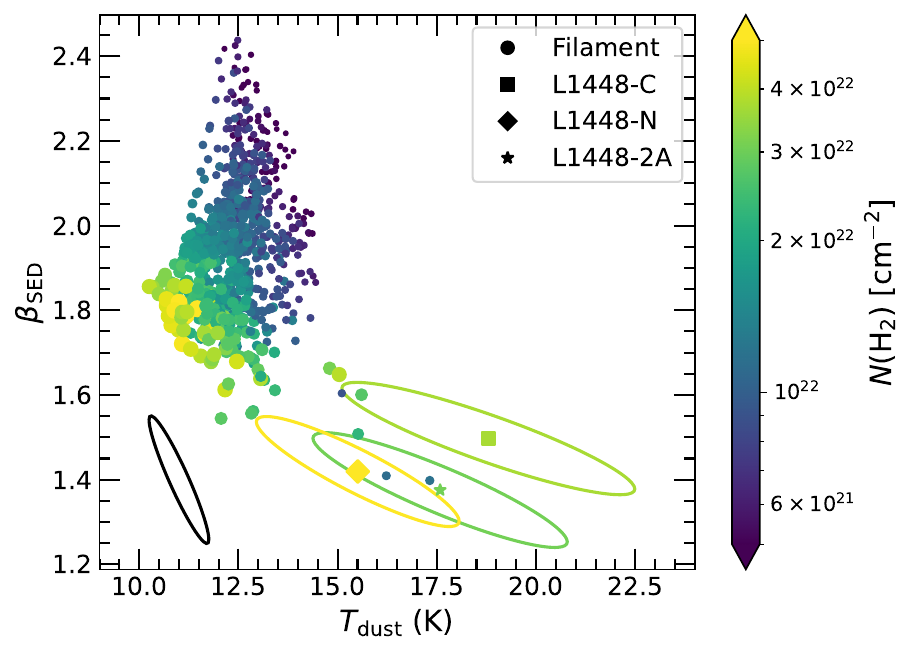}}
  \caption{Dust emissivity index ($\beta_{\mathrm{SED}}$) as a function of dust color temperature ($T_{\mathrm{d}}$) derived from SED fitting for L1448 (see Sec.~\ref{subsec:SED}). Individual protostellar sources are shown as markers, giving the mean value within the footprint defined by \texttt{getsources}. Individual filament pixels are shown as circles. All points are colored by their column density, whose values were taken from the high-resolution HGBS maps \citep{Pezzuto2021} and smoothed to match the angular resolution of our data, as indicated by the colorbar. The marker area also scales linearly with column density. Ellipses show representative $1\sigma$ uncertainties from the covariance posterior of the SED fit at the peak pixel of each source. A single ellipse for the filament points (lower left) and individual ellipses for the three sources. Their orientation reflects the anti-correlation between $T$ and $\beta$ inherent to the fit.}
  \label{fig:correlation_beta_T}
\end{figure}

At large scales, we report the $\beta_{\rm SED}-T_{\rm d}$ relation for the fitted region in Fig.~\ref{fig:correlation_beta_T}. To separate the regions associated with the protostellar cores from that of the surrounding filament, we estimated the source sizes using the \texttt{getsources} algorithm \citep{Men'shchikov2012} applied to the \hersc\ 500~$\mu$m data. This algorithm identifies compact sources by decomposing the emission into spatial scales and constructing a segmentation map that defines the footprint of each source. The footprint encompasses the area of those image pixels that provide a non-negligible contribution to the flux of each source. Since sources have correlated pixels within a beam, we report a single average mean $\beta_{\mathrm{SED}}$ and $T_{\mathrm{dust}}$ within its footprint, while filament values are shown on a pixel-by-pixel basis.

Within the filament, we observe that $\beta_{\rm SED}$ decreases from 2.4 to 1.6 at a nearly constant temperature of approximately $12.5~\rm K$ over an order of magnitude increase in column density. Only when the column density exceeds approximately $3 \times 10^{22}~\mathrm{cm^{-2}}$, an anti-correlation between $\beta_{\rm SED}$ and $T_{\rm dust}$ appears. For the inner envelope, we continue to observe that $\beta$ decreases as the density increases despite only marginal changes in temperature (see Sec.~\ref{subsubsec:Dust_size_evolution}). Given the laboratory experiment described above, changes in temperature alone cannot explain the decrease in $\beta$ observed for temperatures above $15~\rm K$. Taken together, these observations suggest that the variation in $\beta$ is primarily driven by changes in density rather than with temperature.

At smaller spatial scales probed by the interferometric data, we cannot independently assess the $\beta$–$T$ relation, as the temperature profile is computed analytically from the protostellar bolometric luminosity, and not directly measured from data when deriving $\beta$. This limitation points to the need of a higher resolution FIR facility which would allow temperature measurements within the inner protostellar envelopes and would allow to investigate this relationship deeper. Nevertheless, the variations in $\beta$ observed at ALMA scales cannot be explained solely by temperature dependence of the dust emissivity, and likely reflect intrinsic changes in the grain properties.

\subsection{Caveats}
\label{ssec:caveats}

Several effects may affect the $\beta$ profiles measured, whether in the corrections applied to the data, spatial filtering, or emerging from the methods applied. In this section, we revise those factors and explain if they may impact the results or not. 

\subsubsection{The $\beta-T$ degeneracy}
An anti-correlation between the dust temperature and dust emissivity index $\beta$ due to instrumental noise could emerge because both parameters influence the peak and shape of the SED, noise can introduce an artificial anti-correlation between them in the fitted values \citep{Shetty2009a, Shetty2009b, Anderson2012}. While the magnitude of this effect depends on how well the SED is sampled, particularly the peak and the Rayleigh-Jeans tail, it does not introduce a systematic bias in the estimated parameters. Such anti-correlation has been reported in previous observational studies \citep{Dupac2003, Desert2008, Paradis2010, Planck2014}. For L1448, \citet{Chen2016} examined this effect when using data up to $850~\mu$m and concluded that the observed $T-\beta$ trend could not be explained solely by noise. Our data extend further into the Rayleigh-Jeans regime by including observations at $1.15~\rm mm$ and $2\rm~mm$, which further reduces the $T-\beta$ degeneracy when performing an MBB fitting, thus in our data set we expect that the trend show in Fig.~\ref{fig:correlation_beta_T} cannot be explained only by noise.

\subsubsection{Temperature profiles}
Temperature corrections applied at small scales rely on the radial temperature profile described by Eq.~\ref{eq:temp_profile}, which is strictly valid at larger radii in low optical depth regions. A wrong estimation of the $q$ index could introduce biases in the derived $\beta$ values. 

To quantify this effect, we consider a protostar with $4.7~L_{\odot}$, equivalent to L1448-C (see Table~\ref{tab:source_properties}). We explore two cases for the temperature profile exponent, $q=0.2$ and $q=0.4$ (optically thick limit). Using fluxes at 1.3~mm and 3~mm (ALMA Bands 6 and 3), we evaluate the temperature correction term in Eq.~\ref{eq:beta}, which relates the measured spectral index $\alpha$ to $\beta$. We find that at a physical scale of $\sim 4\,000$~au, an incorrect choice of $q$ introduces an offset of only $\sim 0.07$ in $\beta$. At smaller radii, where temperatures are higher, this offset decreases rapidly. Therefore, uncertainties in the $q$-index used in the temperature profile cannot produce a significant offset in $\beta$ across the spatial scales probed. The small magnitude of this effect in $\beta$ indicates that it is unlikely to account for either the low $\beta$ values measured with ALMA or the radial gradients observed. We also note that, for a temperature profile that decreases outward, the correction in Eq.~\ref{eq:beta} can only lower $\beta$ towards smaller radii, so it cannot produce a flat $\beta$ profile.

\subsubsection{Single dish and interferometer comparison}
A further caveat arises from the intrinsic differences between single-dish and interferometric observations. In single-dish observations, measurements probe emission at fixed sky positions and are sensitive to the full column density along the line of sight, whereas interferometric data probes only the emission associated with a given spatial scale, filtering foreground emission at larger scales. Larger scales are expected to contain smaller grains, and thus a higher $\beta$. As a result, measurements from single-dish and interferometric observations may mismatch if the dust population originating from the foreground/background cloud scales, filtered by the interferometer, has a higher $\beta$. This effect could partially contribute to the observed offset we observe for L1448-C in Fig~\ref{fig:beta_profile}. 

Another difference is that our $\beta$ profiles are derived across different wavelength pairs. For NIKA2 between $1.15~\rm mm$ and $2~\rm mm$, and between $1~\rm mm$ and $3~\rm mm$ for ALMA. This matters because the emissivity of amorphous dust does not follow a single power-law opacity, but flattens towards far-IR/sub-mm wavelengths \citep{Strom1977, Mennella1998, Boudet2005, Meny2007, Demyk2022}, so the $\beta$ measured over one frequency range is not necessarily expected to be identical to the value measured over another. Yet, this remains difficult to test robustly, as laboratory measurements typically extend only to $\sim1~\rm mm$ \citep{Demyk2022} and the opacity at longer wavelengths is an extrapolation. If the effect was significant, it would appear as a systematic offset between the NIKA2 and ALMA derived profiles where they overlap. We observe no discontinuity for L1448-2A, suggesting that it is not a dominant contribution. We also stress that, while a frequency dependence of the opacity can shift the two profiles relative to each one, it cannot produce the radial gradient in $\beta$ we observe.

In the visibilities, compact emission can alter the $\beta$ obtained at larger scales by artificially lowering it \citep[see e.g.,][]{Miotello2014, Galametz2019}. This could create differences between the two $\beta$ profiles. By removing the optically thick emission from the visibilities, we mitigate this effect, this allows the interferometric and single-dish measurements to be compared consistently. Because of this correction, we do not expect the difference between the two datasets to be driven by contamination from compact emission.

Differences between the SED and ALMA-derived $\beta$ profiles could also arise from our methods. To test the robustness of our method and evaluate if it could induce a gradient or an offset, we applied the same procedure to synthetic observations. A full description is provided in Appendix \ref{appendix:test_synthetic_obs}. In short, for a model where the emission arises from the same underlying density and temperature distribution at two wavelengths, our method yields consistent values in the derived profiles for visibility-based and image-plane measurements and retrieves the correct $\beta$ used in the dust opacities assumed in the model.

\subsubsection{Filtering and flux calibrations}
In single-dish observations, spatial filtering could artificially induce gradients in $\beta$ if the emission in one of the observed frequency bands was more strongly filtered than the other. Filtering is generally expected for sources bigger than the instantaneous FOV (6.5\arcmin~for NIKA2). To investigate this within our dataset, we characterized the transfer function $T(k)$ of the data reduction pipeline in both NIKA2 bands. We injected noiseless sky models into the data timeline and reduced them exactly as the science data. This test is detailed in Appendix \ref{appendix:filtering}. Our findings indicate that both transfer functions are consistent with unity to within $5\%$ over angular scales of ${36}''\lesssim\theta\lesssim{8}'$ and drops below $0.9$ only beyond $\sim{10}'$, at scales larger than the instrument instantaneous FOV. The two channels are not filtered identically, but the resulting bias on the spectral index is small (see Appendix \ref{appendix:filtering}), well below the changes in $\beta$ we report.  The $\beta_{\rm NIKA2}$ profiles we report extend up to $\sim2$\arcmin~ in scales (see Fig.~\ref{fig:beta_profile}), therefore filtering cannot produce the observed values of $\beta$ nor the radial trend.

In dual-band NIKA2 observations, global flux calibration errors cannot account for the offset observed in $\beta$, since both bands were calibrated using the same calibrator. Any systematic flux offset would cancel out when computing the intensity ratio used to derive the spectral index. Similarly, interferometric measurements can be affected by flux calibration errors, but our dataset shows consistent values for the amplitudes probed by ACA and ALMA data at overlapping $uv$-distances. 

\section{Conclusions}
\label{sec:Conclusions}

We have studied the dust properties from the star-forming filament to the inner envelopes scales in the L1448 Perseus clump, by combining multi-band measurements from large-scale ($30\,000~\rm au$) using single-dish NIKA2 data, down to small scale (400 au) using interferometric visibilities obtained from ALMA. This approach allows us to trace the variation of the dust emissivity index across almost two orders of magnitude, filling the gap that usually divides single-dish and interferometric measurements. From this multiscale approach, we derived radial profiles of the dust emissivity index.

Our results can be summarized as follow:
\begin{enumerate}
    \item There is a systematic and monotonic evolution of the dust emissivity index, $\beta$, with decreasing spatial scale from the relatively homogeneous values consistent with ISM measurements at cloud scales ($\beta \sim 2$) to much smaller values measured inside the inner envelope ($\beta \sim0.3$ at scales of $\sim400~\rm au$).
    \item Thanks to the multiscale nature of our study, we conclude that the decrease in $\beta$ primarily correlates with increasing gas density rather than temperature, over the probed ranges of $n_{\mathrm{H_{2}}}\sim 10^{5} -  10^{7}~\mathrm{cm^{-3}}$ and $T\sim10-37~\rm K$. Both L1448-2A and L1448-C, which originate from the same initial reservoir but differ in luminosity while having similar envelope masses, show very similar $\beta$ evolution with increasing density.
    \item We explore the physical process that potentially drive the variations of $\beta$. The values of $\beta$ measured in the inner envelopes cannot be explained by variations in grain composition, structure, ice mantles, nor optical depth effects. Our data shows a dependency of $\beta$ with density $n^{-0.17}$. In addition, we find a tentative drop with $\beta$ dropping more steeply once $n_{\mathrm{H_{2}}}\sim2\times10^{6}~\mathrm{cm^{-3}}$, which suggests a change of the dust grain sizes with density could drive the observed correlation. While this result is at odds with current models of dust evolution, they call for further investigations and better physical models for dust coagulation in the dense gas, long before the material incorporates the disk.
\end{enumerate}

\begin{acknowledgements}
The authors thank Nathalie Ysard for useful discussions on the topic and paper early analysis. This project has received funding from the European Research Council (ERC) under the European Union Horizon 2020 research and innovation programme (PEBBLES project, grant agreement N$\degr$ 101098309, PI: A.J. Maury). This work was also partly supported by the Spanish program Unidad de Excelencia María de Maeztu CEX2020-001058-M, financed by MCIN/AEI/10.13039/501100011033, and by the MaX-CSIC Excellence Award MaX4-SOMMA-ICE. This work is part of the first author's doctoral thesis, within the framework of the Doctoral Program in Physics at the Universitat Autònoma de Barcelona. A.Z. and L.C. acknowledge support from the Joint ALMA Observatory Visitor Program. The NIKA2 data were processed using the Pointing and Imaging In Continuum (PIIC) software, developed by Robert Zylka and Stefano Berta at the Institut de Radioastronomie Millimetrique (IRAM) and distributed by IRAM via the GILDAS pages. PIIC is the extension of the MOPSIC data reduction software to the case of NIKA2 data. This research has made use of data from the \hersc\ Gould Belt survey (HGBS) project (http://gouldbelt-herschel.cea.fr). The HGBS is a \hersc\ Key Programme jointly carried out by SPIRE Specialist Astronomy Group 3 (SAG 3), scientists of several institutes in the PACS Consortium (CEA Saclay, INAF-IFSI Rome and INAF-Arcetri, KU Leuven, MPIA Heidelberg), and scientists of the \hersc\ Science Center (HSC). This paper makes use of the following ALMA data: ADS/JAO.ALMA$\#$2015.1.01194.S, ADS/JAO.ALMA$\#$2016.1.00604.S, ADS/JAO.ALMA$\#$2017.1.00053.S, ADS/JAO.ALMA$\#$2017.1.01693.S, ADS/JAO.ALMA$\#$2018.1.00827.S, ADS/JAO.ALMA$\#$2018.1.00934.S, ADS/JAO.ALMA$\#$2019.2.00167.S, ADS/JAO.ALMA$\#$2021.1.00276.S. ALMA is a partnership of ESO (representing its member states), NSF (USA) and NINS (Japan), together with NRC (Canada), NSTC and ASIAA (Taiwan), and KASI (Republic of Korea), in cooperation with the Republic of Chile. The Joint ALMA Observatory is operated by ESO, AUI/NRAO and NAOJ.
\end{acknowledgements}
 
\vspace{-10pt}
\bibliographystyle{aa.bst}
\bibliography{mybiblio_MY.bib}

\clearpage
\newpage
\begin{appendix}
\nolinenumbers

\section{NIKA2 map making}
\label{appendix:NIKA2}

Data were reduced using the \texttt{PIIC} software\footnote{The Pointing and Imaging In Continuum (PIIC), publicly available at \url{https://publicwiki.iram.es/PIIC/}} \citep{Zylka2013, piic19} and the customized \texttt{mapTP\_a2\_wc\_template.piic} pipeline provided by the package. This pipeline includes steps of flat field correction, correlation correction, sky subtraction, baseline subtraction, calibration, geometry association and map re-gridding. The final maps are about 0.5$^{\circ}\times0.5^{\circ}$. The pointing was regularly checked, using \texttt{0316+413} as a reference source. Corrections remained below 4\arcsec\ for all scans. 

Diffuse emission around the bright sources can be recovered using an iterative reduction method. First, one produces the 0th iteration of the map, to define the footprint of for emission (source model) to optimize our source definition before proceeding  with this iterative procedure. From the first iteration, we produced a source model by keeping data with a signal-to-noise ratio (S/N) level above 3. For each iteration, the source definition  was refined using the same S/N criterion. The pipeline was then re-run using this source model. Our 1.15 and 2mm maps are finally converted in MJy~sr$^{-1}$. We note that the pixel resolution of the maps is 3\arcsec, close to the Nyquist sampling limit of the 1~mm observation  ($\lambda$/2D $\sim$ 4.3\arcsec, with D=27.5m the diameter of the IRAM 30m primary mirror).

\section{ALMA observations}

We summarize here the ALMA datasets used in this work in Table~\ref{tab:ALMA_observations}. The archival material comprises both high-resolution observations that trace the compact structures, and lower-resolution, high-sensitivity data that recover the extended envelopes. We used a combination of ALMA 12~m and ACA datasets to ensure a multiscale coverage. For completeness, we also present the continuum emission maps obtained from the combination of ACA and ALMA 12~m data for each source.

\renewcommand{\arraystretch}{1.2}
\begin{table*}
\caption{Summary of ALMA observations for L1448-C, L1448-2A, and L1448-N}
    \centering
    \begin{tabular}{c c c c c c c} 
    \hline\hline
    System & Project code & PI: & Integration (s) & Resolution & Baseline (k$\lambda$) & Frequency (GHz)\\
    \hline
    \multicolumn{7}{c}{\textbf{Band 3}}\\
    \hline
    L1448-C & 2018.1.00827.S & Cox, E. & 2358.7 & 0\arcsec.3 & 7-1105 & 91-103\\
    L1448-C & 2019.2.00167.S & Galametz, M. & 2806.3 & 14\arcsec.6 & 2-13 & 91-103\\
    L1448-2A & 2017.1.01693.S & Hsieh, T. & 393.1 & 1\arcsec.3 & 4-232 & 93-95\\
    L1448-2A & 2019.2.00167.S & Galametz, M. & 2806.3 & 14\arcsec.6 & 2-13 & 91-103\\
    \hline
    \multicolumn{7}{c}{\textbf{Band 4}}\\
    \hline
    L1448-N & 2019.2.00167.S & Galametz, M. & 6350.4 & 9\arcsec.1 & 3-19 & 139-150\\
    L1448-N & 2021.1.00276.S & Looney, L. & 4173 & 0\arcsec.2 & 63-5620 & 137-152\\
    \hline
    \multicolumn{7}{c}{\textbf{Band 6}}\\
    \hline
    L1448-C & 2015.1.01194.S & Tafalla, M. & 2589.0 & 0\arcsec.4 & 10-789 & 216-231\\
    L1448-C & 2019.2.00167.S & Galametz, M. & 818.5  & 5\arcsec.7 & 6-29 & 223-243\\
    L1448-2A & 2016.1.00604.S & Kwon, M. & 6168.9 & 0\arcsec.4 & 13-822 & 225-241\\
    L1448-2A & 2019.2.00167.S & Galametz, M. & 818.5 & 5\arcsec.7 & 10-28 & 225-241\\
    L1448-N & 2017.1.00053.S & Tobin, J. & 1058 & 0\arcsec.2 & 19-3279 & 218-234\\
    L1448-N & 2018.1.00934.S & Murillo, N. & 3024 & 5\arcsec.8 & 6-33 & 232-234\\
    \hline
\end{tabular}
\label{tab:ALMA_observations}
\end{table*}

\subsection{Data reduction and self-calibration}
\label{appendix:ALMA_data_reduction}

The delivery and calibration of the raw visibility data was performed by the European ALMA Regional Centre (ARC) staff using the Common Astronomy Software Applications \citep[\texttt{CASA}, ][]{McMullin2007} pipeline. Different pipeline versions were employed depending on the specific project (see Table \ref{tab:ALMA_observations}).

We inspected the metadata of each measurement set (MS) using the \texttt{listobs} task and subdivided the MS with the \texttt{split} task to extract a dataset containing only the continuum spectral windows of the science target. For datasets obtained as mosaics, we selected only data from the source the field is centered on. We examined individual ALMA cubes for each execution block (EB) and performed additional flagging where line emission was identified. To create a pseudo-continuum dataset with reduced size and improved signal-to-noise ratio (SNR), we spectrally averaged the continuum channels, ensuring the number of averaged channels avoided bandwidth smearing \citep{Brindle1999}. 

We then imaged each dataset using \texttt{tclean}, and visualized the resulting maps with the Cube Analysis and Rendering Tool for Astronomy \citep[CARTA,][]{Comrie2021}. To determine the emission peak position, we performed a Gaussian fitting. For datasets with sufficient angular resolution to resolve multiple components, we restricted the fitting region to the brightest source: L1448-2Aa, L1448-C(N), and L1448 IRS3B-C, respectively. Using the \texttt{phaseshift}, we shifted each EB phase center to the measured emission peak position. Once all EBs were aligned, for each source we set a common coordinate in the metadata using \texttt{fixplanets} and concatenated EBs belonging to the same Member Observing Unit Set (MOUS).

We then performed phase self-calibration on the aligned pseudo-continuum MOUS visibilities individually. We applied progressively shorter phase solution intervals, \texttt{solint = `inf', `60s', `10s', `int'}, stopping when the signal-to-noise ratio (SNR) stopped to improve significantly. The reference antenna was chosen based on its position within the array. To enhance the SNR of the solutions, we set \texttt{gaincal = `T'}. Solutions tables were applied to the aligned pseudo-continuum visibilities using the \texttt{gaincal} task, with \texttt{applymode = `calonly'} to ensure that only calibration was applied without flagging additional data. After each iteration, we re-imaged the data using \texttt{tclean} with a Briggs weighting scheme (\texttt{robust = 0.5}) and performed interactive masking. The SNR and source flux were compared to previous iterations to determine whether to proceed with the next solution interval. The self-calibration process resulted in improvements between 300\% to 900\% in the peak SNR. We did not attempt amplitude self-calibration. Once all MOUS where self-calibrated, we combined them.

\begin{figure}
  \centering
  \includegraphics[width=\hsize,height=0.3\textheight,keepaspectratio]{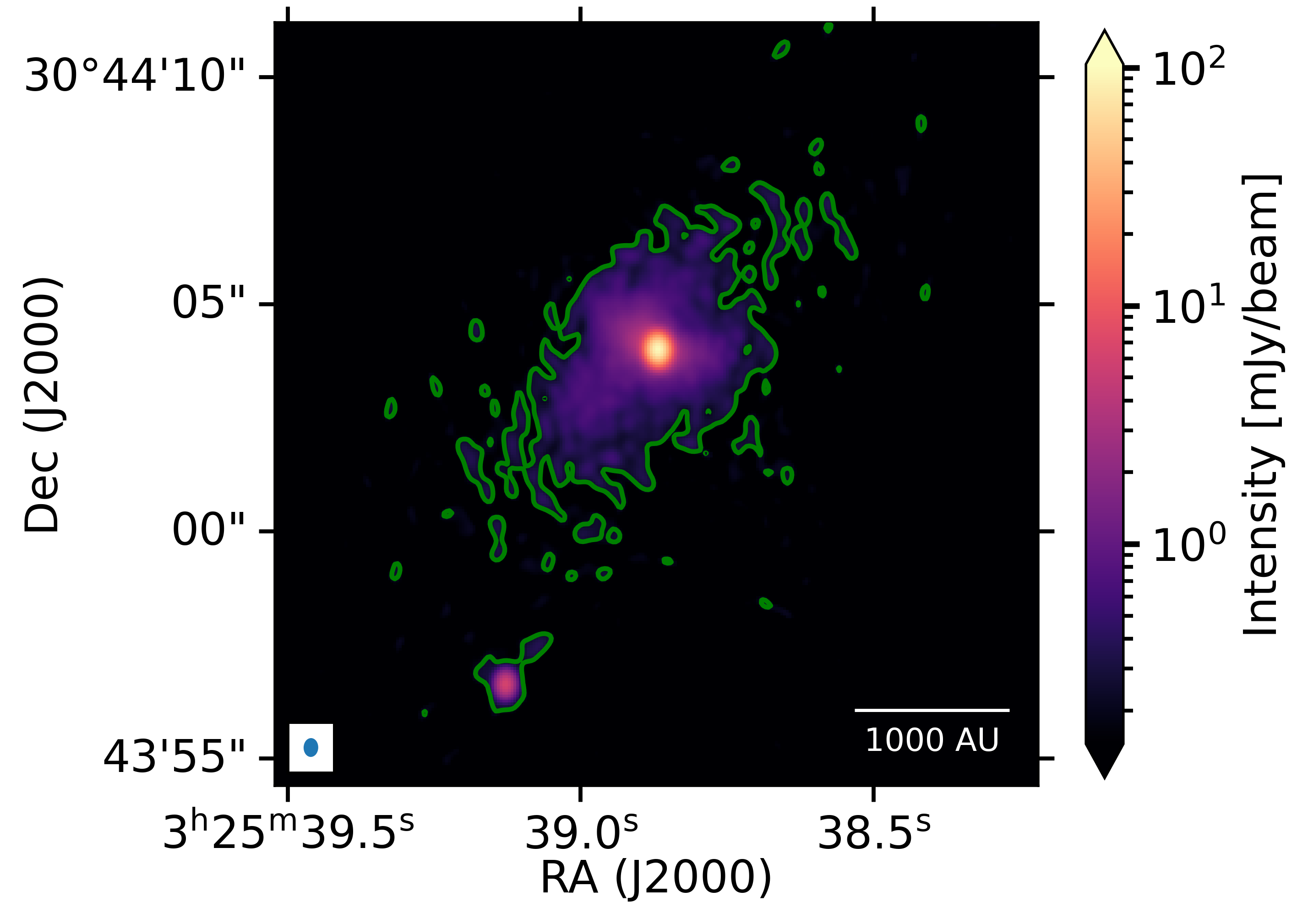}\par\smallskip
  \includegraphics[width=\hsize,height=0.3\textheight,keepaspectratio]{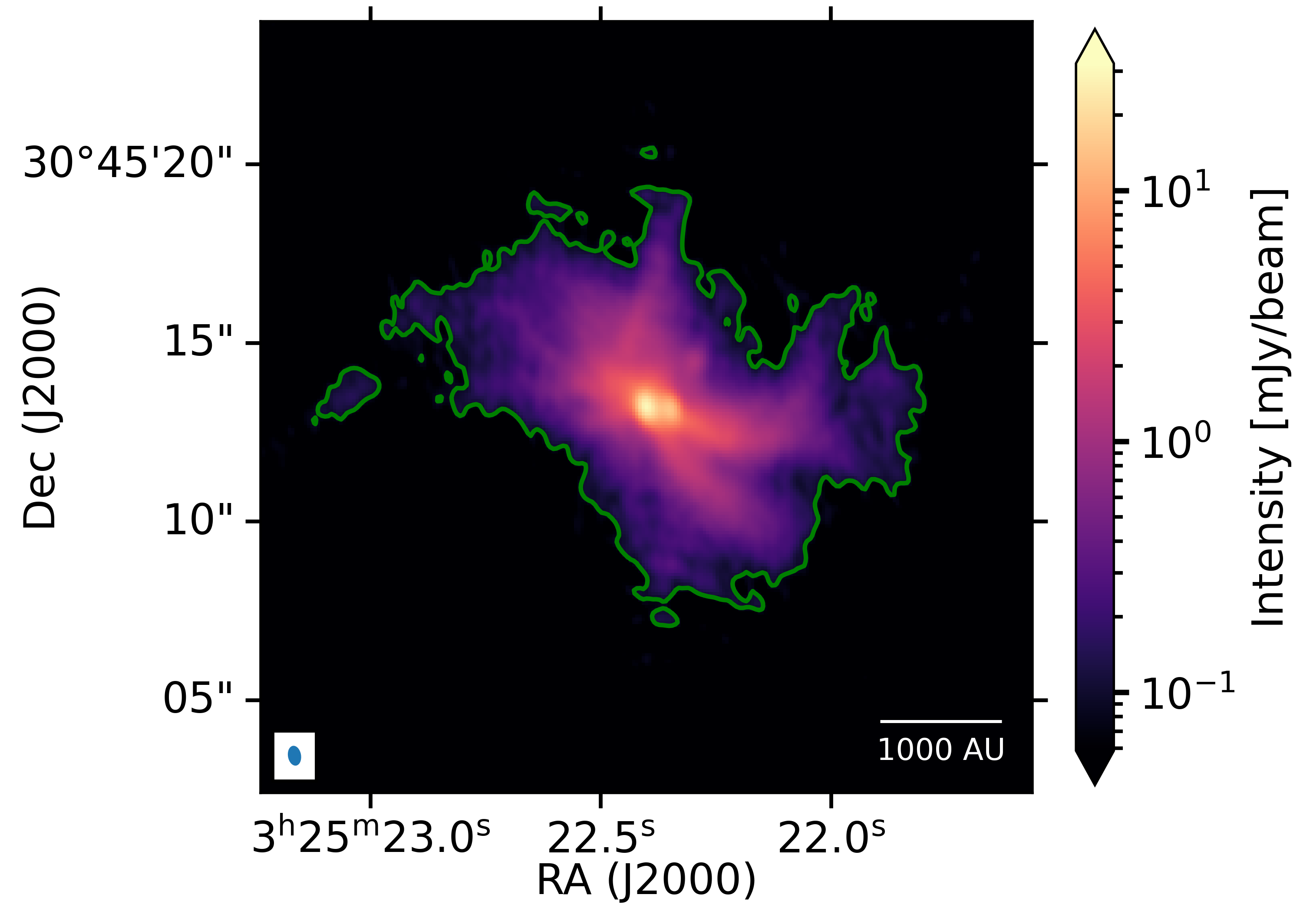}\par\smallskip
  \includegraphics[width=\hsize,height=0.3\textheight,keepaspectratio]{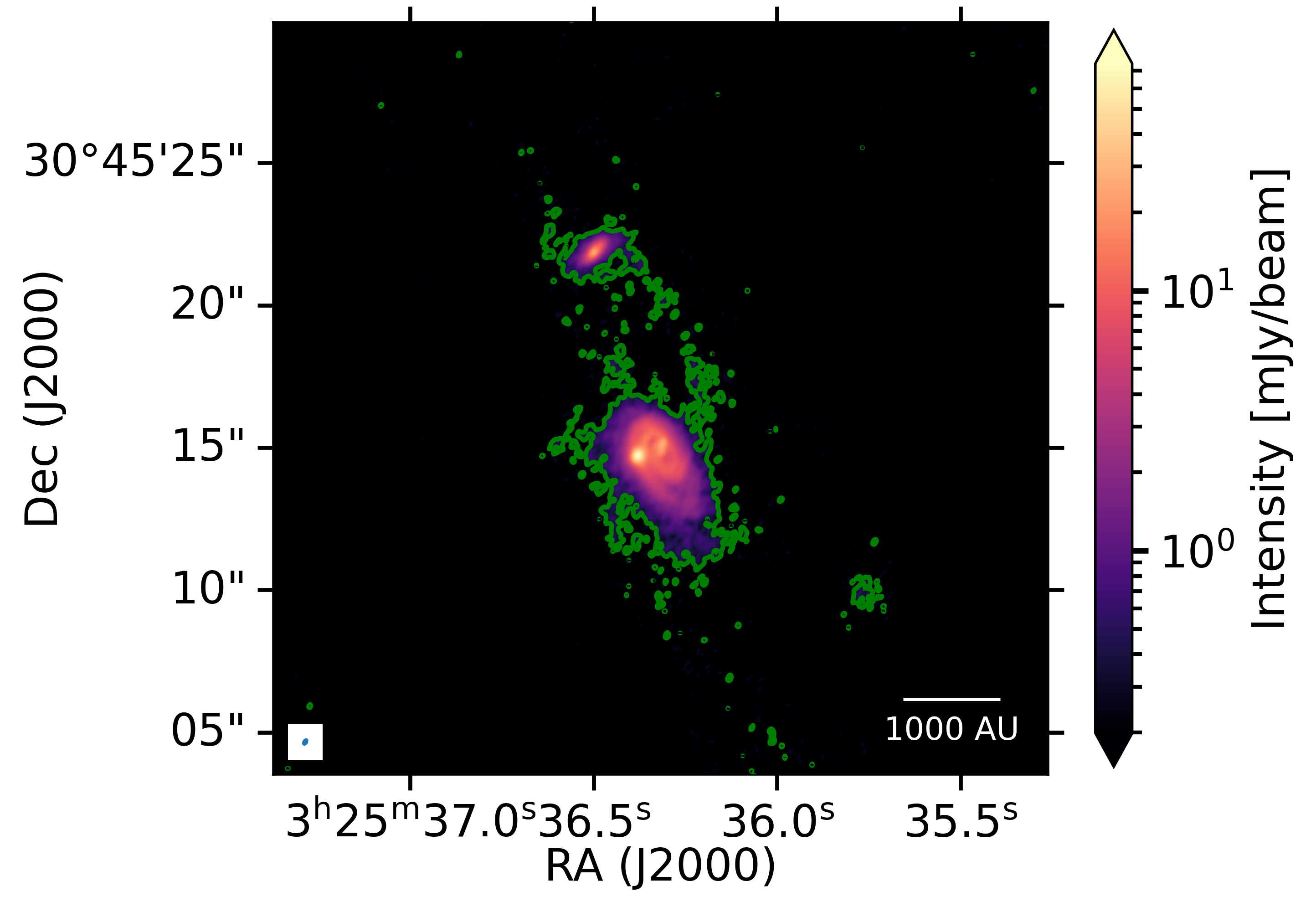}
  \caption{Dust continuum emission maps of the combined 7m + 12m ALMA data. From top to bottom: L1448-C (Band 6), L1448-2A (Band 6), and L1448-N (Band 4). The synthesized beam is shown in the lower left corner of each panel as an ellipse. The green contour shows the $3\sigma$ level. In the lower right corner there is a scale bar showing the equivalent of $1\,000~\rm au$.}
  \label{fig:ALMA_continuum}
\end{figure}

\subsection{Visibility profiles}
\label{appendix:visibility_profiles}

\begin{figure}
  \centering
  \includegraphics[width=\hsize,height=0.3\textheight,keepaspectratio]{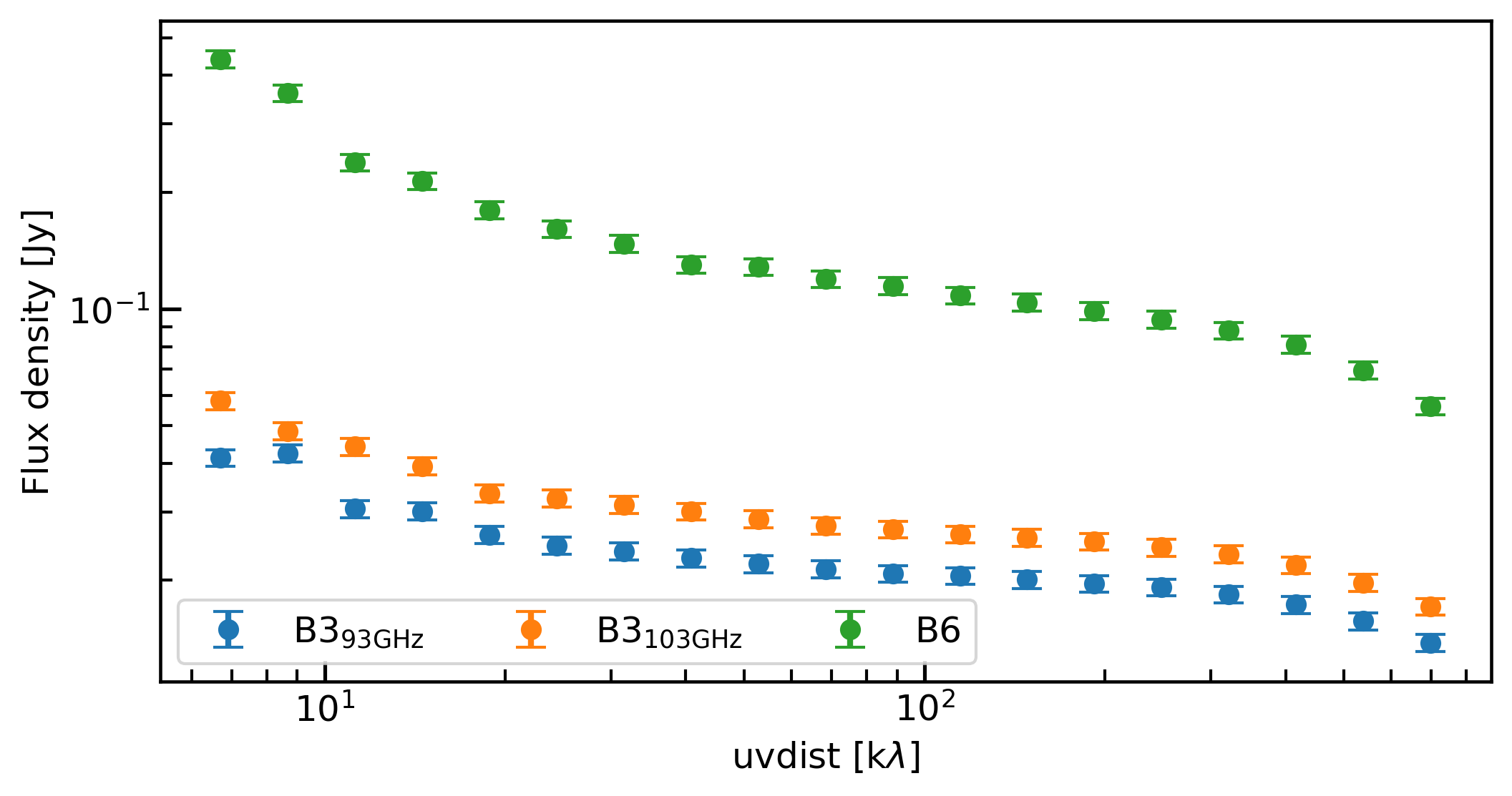}\par\smallskip
  \includegraphics[width=\hsize,height=0.3\textheight,keepaspectratio]{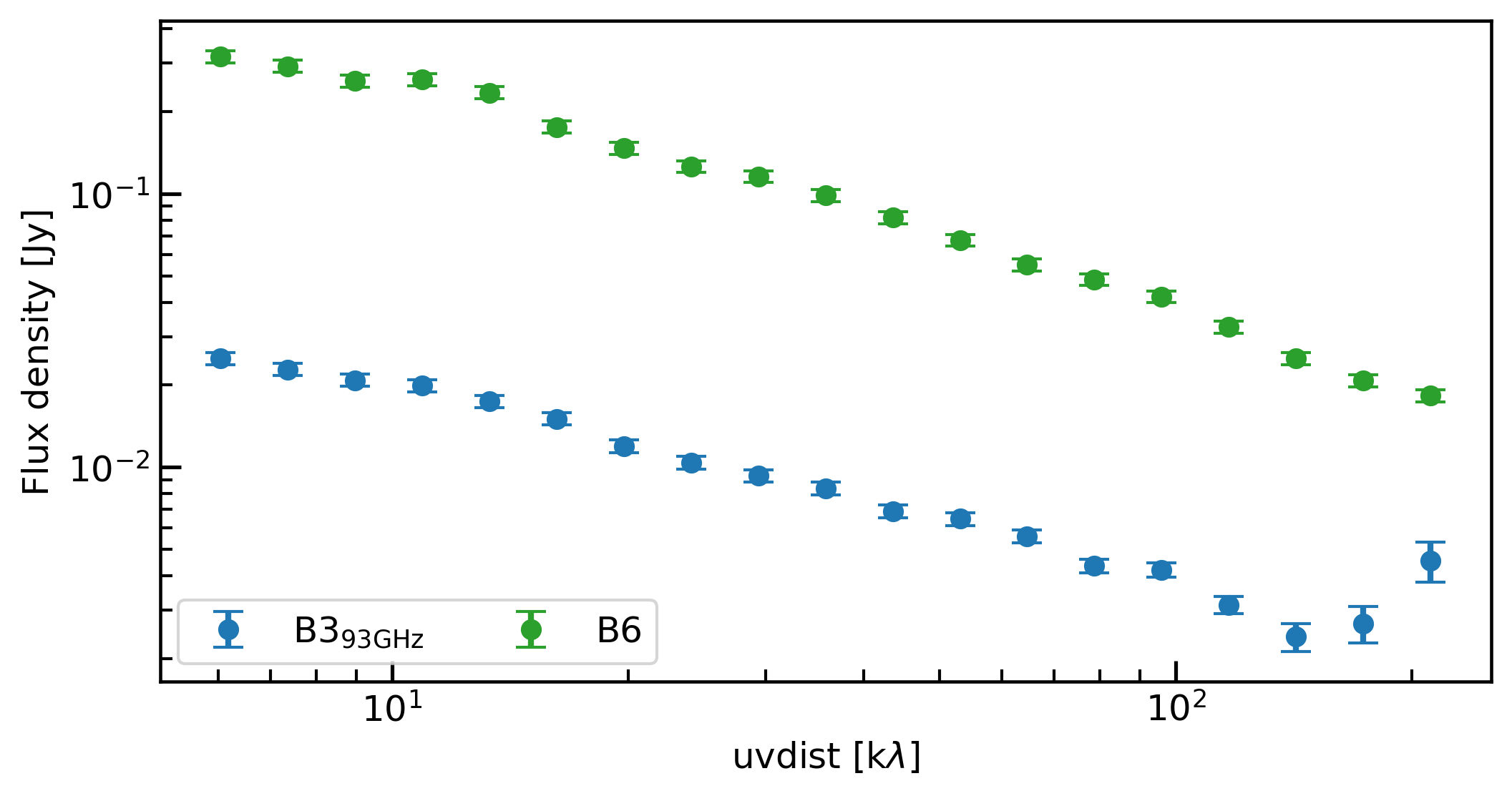}\par\smallskip
  \includegraphics[width=\hsize,height=0.3\textheight,keepaspectratio]{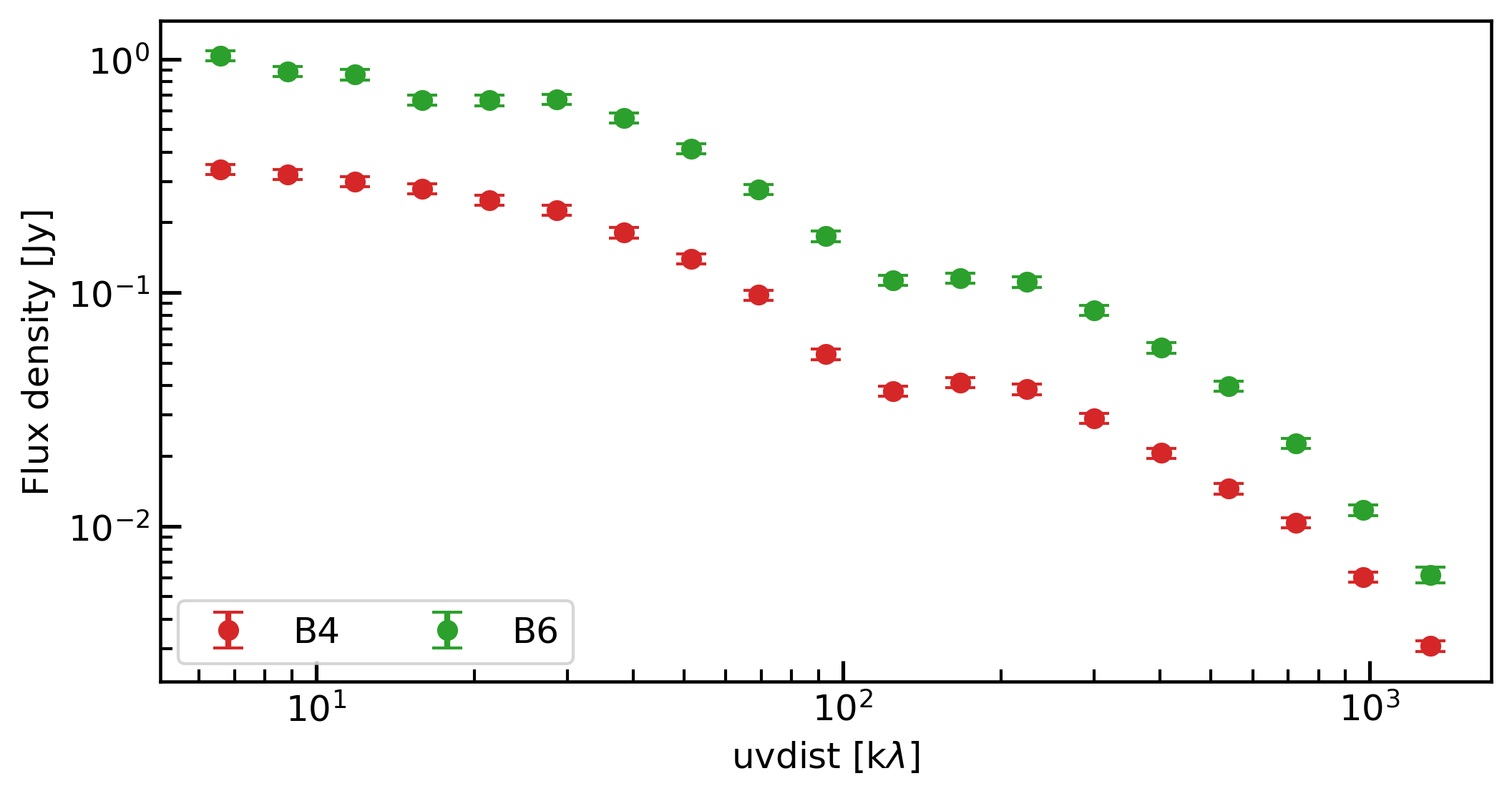}
  \caption{Binned azimuthally averaged amplitude profiles as a function of uv-distance. From top to bottom: L1448-C, L1448-2A, and L1448-N. The colors of the markers indicate which band they belong to.}
  \label{fig:vis_profile}
\end{figure}

Fig.~\ref{fig:vis_profile} shows binned visibility amplitudes profiles as function of uv-distance, for each source at each available ALMA Band.

\section{Effect of the $70~\mu$m point on the SED fitting}
\label{appendix:example_SED_fitting}

\begin{figure}
  \resizebox{\hsize}{!}{\includegraphics{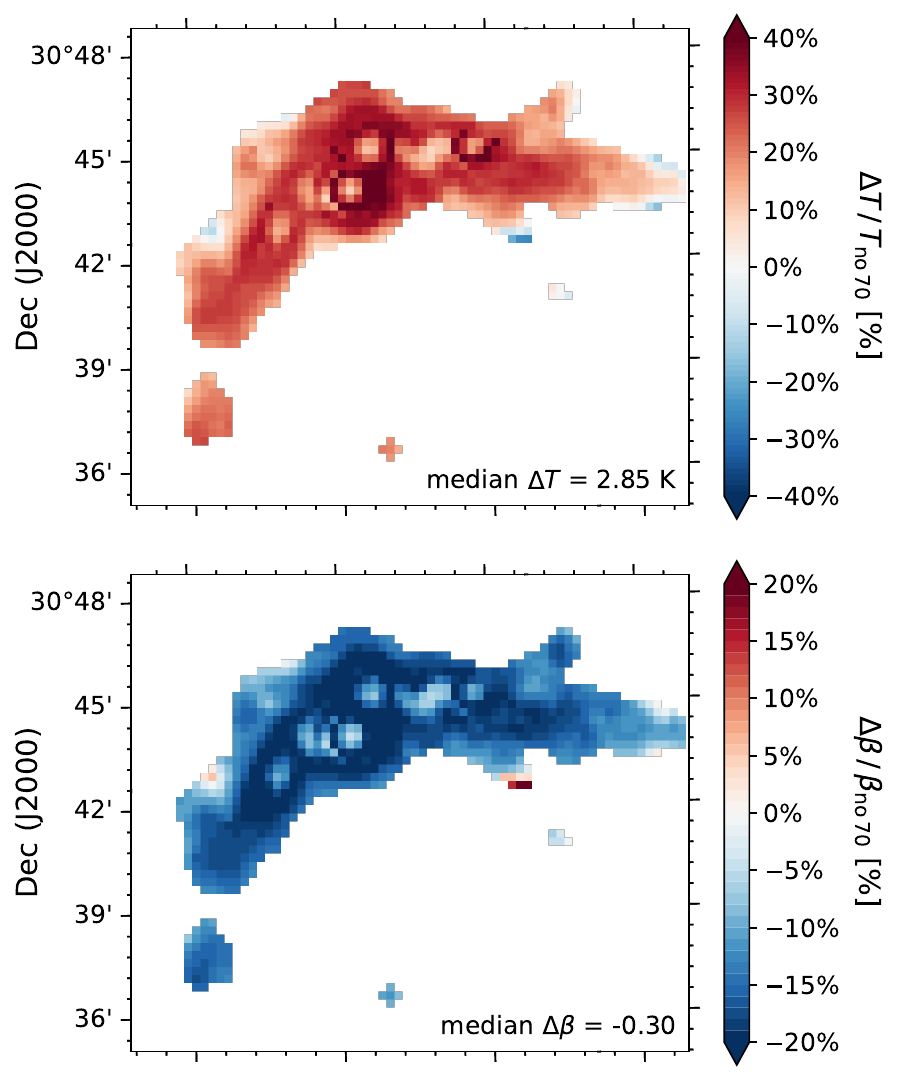}}
  \caption{Difference in the quantities derived by the SED fitting when including or excluding the $70~\mu\rm m$ point. Top: Temperature map. Bottom: Dust emissivity map. The median difference is given in the lower right corner of each panel.}
  \label{fig:L1448_T_beta_difference_maps}
\end{figure}

Excluding the $70~\mu$m data point is the standard procedure in the Herschel Gould Belt Survey papers \citep[see e.g.][]{Pezzuto2021}. The reason is that this band receives a significant contribution from small grains \citep{Schnee2008}, which are not in thermal equilibrium and therefore cannot be described by a single modified black-body. Here we quantify the effect of this choice. We repeated the SED fitting described in Sec.~\ref{subsec:SED} both including and excluding the $70~\mu$m point. Fig.~\ref{fig:L1448_T_beta_difference_maps} shows the resulting difference in the temperature and $\beta$ maps.

\begin{figure*}
  \includegraphics[width=17cm]{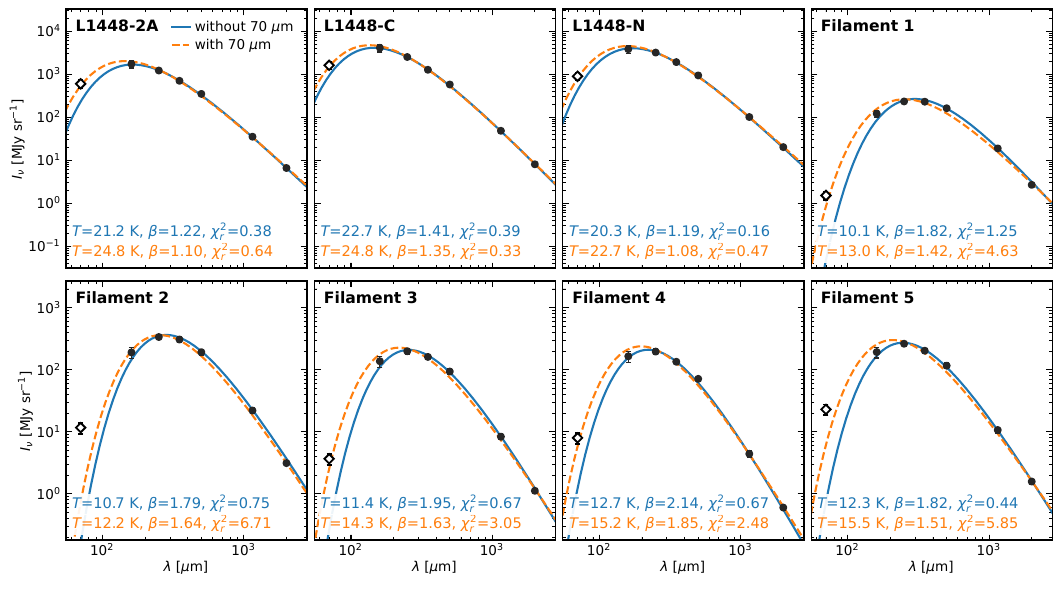}
    \caption{Spectral energy distributions at eight characteristic locations in L1448, fitted with and without the 70~$\mu$m point. Filled circles are the \textit{Herschel} (160, 250, 350 and 500\,$\mu$m) and NIKA2 (1.15 and 2\,mm) photometry used in both fits. The open diamond is the \textit{Herschel}/PACS 70~$\mu$m measurement. Error bars are adopted from the calibration uncertainties (see Sec.~\ref{subsec:SED}). Blue solid line is the fit to the six bands from 160\,$\mu$m to 2\,mm, and the orange dashed line is the fit that also includes 70~$\mu$m. The best-fit temperature, dust emissivity index and reduced $\chi^2$are listed in each panel.}
  \label{fig:L1448_SED_70mum_comparison}
\end{figure*}

Including the $70~\mu\rm m$ data yields higher temperatures, by a median of $\sim2.9~\rm K$, corresponding to an increase of roughly $20\%$, and yields lower $\beta_{\rm SED}$, by a median of 0.3. The increment in temperature is more prominent towards region surrounding the protostars, while the decrement in $\beta_{\rm SED}$ is observed towards the filamentary regions. The fits are also poorer, the median reduced $\chi^{2}$ is $3.33$ when the $70~\mu$m is included, against $0.65$ when it is excluded. This effect is illustrated in Fig.~\ref{fig:L1448_SED_70mum_comparison}, were we show the fit including or excluding the 70~$\mu$m point for a set of representative points in L1448. The offset of $0.3$ in $\beta_{\rm SED}$ applies to the emissivity index derived directly from the SED fit. Throughout the manuscript, however, the SED fitting is only used to obtain a temperature map, which then converts the spectral index measured from the millimeter data into $\beta_{\rm NIKA2}$. The $70~\mu\rm m$ choice could only propagate into our results through $T$. At $20~\rm K$ an increment of roughly $\sim20\%$ on the derived temperatures corresponds  to a shift of $\sigma_{\beta} = 0.056$ (see Appendix~\ref{appendix:uncertainty_propagation}). Our results are therefore unaffected by this choice.

\section{Comparison with HGBS temperature map}
\label{appendix:temperature_comparison}

We compare our dust temperature map derived from SED fitting with the temperature estimates presented by \citet{Pezzuto2021}. In that work, temperatures were obtained from an MBB fitting (Eq.~\ref{eq:MBB}) using only \hersc\ $160 - 500~\mu$m data, assuming a fixed emissivity index $\beta = 2$, and without applying color corrections. In our analysis, we extend the fitted wavelength range to $160~\mu\mathrm{m} - 2~\mathrm{mm}$, treat $\beta$ as a free parameter and include color corrections(see Sec.~\ref{subsec:SED}).

\begin{figure}
  \resizebox{\hsize}{!}{\includegraphics{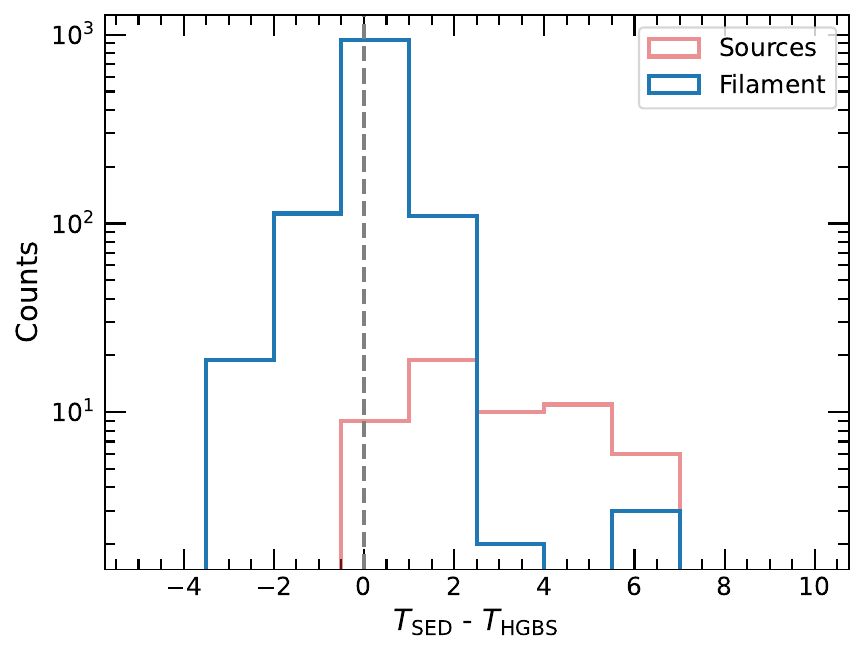}}
  \caption{Histogram of the temperature difference between our SED-derived temperature map, $T_{\mathrm{\rm d}}$, and the publicly available HGBS temperature map, $T_{\mathrm{HGBS}}$. Filament pixels are shown in blue, while protostellar source pixels are shown in red.}
  \label{fig:temperature_diff_histogram}
\end{figure}

In Fig~\ref{fig:temperature_diff_histogram}, we show the distribution of temperature differences between the two  maps, separating filament from protostellar source pixels. Filament temperatures are broadly distributed around $\Delta T = 0$, indicating good agreement. Temperature towards the protostellar sources are systematically higher in our maps. This temperature offset arises from allowing $\beta$ to vary in the SED fitting. In these regions values of $\beta < 2$ are inferred, therefore fixing $\beta = 2$ leads to lower temperature estimates in the HGBS maps.

As a consistency check, we repeated our SED fitting following the same assumptions as \citet{Pezzuto2021}, restricting the fit to $160 - 500~\mu$m data and fixing $\beta = 2$. Under these conditions, the resulting temperatures agree within $<1$~K for all pixels.

\section{Propagation of uncertainties on $\alpha$ and $\beta$}
\label{appendix:uncertainty_propagation}

In this appendix, we detail the propagation of uncertainties on both $\alpha$ and $\beta$, assuming that they are evaluated across a single pair of frequencies. The uncertainty associated with $\alpha$ is given by:

\begin{equation}
    \sigma^{2}_{\alpha} =
    \left( \frac{\partial \alpha}{\partial I_{1}} \right)^{2}\sigma_{I_{1}}^{2}
    + \left( \frac{\partial \alpha}{\partial I_{2}} \right)^{2}\sigma_{I_{2}}^{2},
    \label{eq:uncertainty_alpha}
\end{equation}

where $I_{1}$ and $I_{2}$ are the intensities at $\nu_{1}$ and $\nu_{2}$, respectively, and where the two measurements are assumed to be independent. Computing the partial derivatives of Eq.~\ref{eq:alpha}and plugging them into Eq.~\ref{eq:uncertainty_alpha} results in:

\begin{equation}
    \sigma_{\alpha}^{2} = \left( \frac{1}{\ln{(\nu_{1}/\nu_{2})}} \right)^{2} \left( \frac{\sigma_{I_{1}}^{2}}{I_{1}^{2}} + \frac{\sigma_{I_{2}}^{2}}{I_{2}^{2}} \right),
    \label{eq:sigma_alpha}
\end{equation}

where $\sigma_{I_{1}}$ and $\sigma_{I_{2}}$ are the uncertainties on the intensities, including the flux calibration errors. 

To derive the uncertainty on $\beta$ it is convenient to express the second term of Eq.~\ref{eq:beta} as a two-point slope instead of a derivative, consistenly with the way $\alpha$ is measured. This reads as:

\begin{equation}
     \beta = \alpha - \frac{1}{\ln{(\nu_{1}/\nu_{2})}} \ln{\frac{B_{\nu_{1}}(T_{\rm d})}{B_{\nu_{2}}(T_{\rm d})}},
    \label{eq:beta_2point}
\end{equation}

which is exact for any frequency separation and reduces to Eq.~\ref{eq:beta} in the limit $\nu_{1} \rightarrow \nu_{2}$. To make the second terms more manageable, we define the dimensionless frequency $x_{i} \equiv h\nu_{i}/(kT_{d})$ and write the logarithm of the Planck function as: 
\begin{equation}
    \ln{B_{\nu_{i}}(T_{\rm d})} = \ln{\frac{2h}{c^{2}}} + 3\ln{\nu_{i}}
    - \ln{(e^{x_{i}}-1)}.
    \label{eq:planck_func_log}
\end{equation}

Subtracting Eq.~\ref{eq:planck_func_log} between the two bands, the constant
term cancels and the second term of Eq.~\ref{eq:beta_2point} becomes
\begin{equation}
    \Lambda \equiv \frac{1}{\ln{(\nu_{1}/\nu_{2})}}
    \ln{\frac{B_{\nu_{1}}(T_{\rm d})}{B_{\nu_{2}}(T_{\rm d})}}
    = 3 - \frac{1}{\ln{(\nu_{1}/\nu_{2})}}
    \ln{\frac{e^{x_{1}}-1}{e^{x_{2}}-1}}.
    \label{eq:lambda}
\end{equation}

We then differentiate Eq.~\ref{eq:planck_func_log} for a single band. The dust temperature enters only through $x_{i}$, with $\partial x_{i} / \partial T_{\rm d} = -x_{i}/T_{\rm d}$, so that
\begin{equation}
    \frac{\partial\ln{B_{\nu_{i}}}}{\partial T_{\rm d}} = -\frac{e^{x_{i}}}{e^{x_{i}}-1}\frac{\partial x_{i}}{\partial T_{\rm d}} = \frac{g(x_{i})}{T_{\rm d}}, \qquad g(x) \equiv \frac{xe^{x}}{e^{x}-1}.
    \label{eq:dlnB_dT}
\end{equation}

Because $\Lambda$ is simply the difference of two such logarithms divided by the constant $\ln{(\nu_{1}/\nu_{2})}$, applying Eq.~\ref{eq:dlnB_dT} to each band gives

\begin{equation}
    \frac{\partial\Lambda}{\partial T_{\rm d}} = \frac{g(x_{1}) - g(x_{2})}{\ln{(\nu_{1}/\nu_{2})}\,T_{\rm d}}.
    \label{eq:dlambda_dT}
\end{equation}

Finally, the error on $\beta$ depends on two inputs, the measured intensity ratio and the adopted dust temperature. Propagating both errors, and assuming them to be independent,

\begin{equation}
    \sigma_{\beta}^{2} = \sigma_{\alpha}^{2} + \left( \frac{\partial \Lambda}{\partial T_{\rm d}} \right)^{2}\sigma_{T_{\rm d}}^{2}.
    \label{eq:sigma_beta}
\end{equation}

\section{Galario fit results}

Table~\ref{tab:galario_results} present best-fitting results from \texttt{galario} for L1448-C, using a model consisting of an envelope and compact component. The quoted uncertainties are statistical only, derived from the 16th and 84th percentiles of the MCMC marginalized distribution.

\begin{table*}[ht]
\caption{Best-fit parameters for ALMA visibility fitting obtained with \texttt{galario} for L1448-C.}
\label{tab:galario_results}
\centering
\footnotesize
\setlength{\tabcolsep}{5pt}
\renewcommand{\arraystretch}{1.5}
\begin{tabular}{c c c c c c c c c c c} 
    \hline\hline
         Band & $f_{0}$ & $\sigma$ & inc & PA & $f_{1}$ & $R_{\rm in}$ & $R_{\rm out}$ & p & dRA & dDec \\
     & [Jy~sr$^{-1}$] & [\arcsec] & [deg] & [deg] & [Jy~sr$^{-1}$] & [\arcsec] & [\arcsec] &  & [\arcsec] & [\arcsec] \\
    (1) & (2) & (3) & (4) & (5) & (6) & (7) & (8) & (9) & (10) & (11) \\
    \hline
    3 (90 GHz)  & 11.00$^{+0.01}_{-0.01}$ & 0.04$^{+3\mathrm{E}-4}_{-3\mathrm{E}-4}$ & 46.07$^{+1.06}_{-1.09}$ & 66.39$^{+1.05}_{-1.06}$ & 7.92$^{+0.02}_{-0.02}$ & 0.44$^{+0.02}_{-0.02}$ & 12.41$^{+0.14}_{-0.07}$ & 2.21$^{+0.01}_{-0.01}$ & -0.01$^{+1\mathrm{E}-4}_{-1\mathrm{E}-4}$ & 0.02$^{+1\mathrm{E}-4}_{-1\mathrm{E}-4}$ \\
    3 (103 GHz) & 11.07$^{+5\mathrm{E}-3}_{-5\mathrm{E}-3}$ & 0.05$^{+1\mathrm{E}-4}_{-1\mathrm{E}-4}$ & 45.50$^{+0.7}_{-0.7}$ & 69.04$^{+0.7}_{-0.7}$ & 8.06$^{+4\mathrm{E}-3}_{-4\mathrm{E}-3}$ & 0.50$^{+2\mathrm{E}-3}_{-2\mathrm{E}-3}$ & 10.50$^{+0.05}_{-0.05}$ & 2.36$^{+6\mathrm{E}-3}_{-6\mathrm{E}-3}$ & 0.00$^{+1\mathrm{E}-4}_{-1\mathrm{E}-4}$ & 0.02$^{+1\mathrm{E}-4}_{-1\mathrm{E}-4}$ \\
    6          & 11.54$^{+2\mathrm{E}-3}_{-2\mathrm{E}-3}$ & 0.06$^{+1\mathrm{E}-4}_{-1\mathrm{E}-4}$ & 54.67$^{+0.14}_{-0.14}$ & 66.90$^{+0.12}_{-0.13}$ & 8.95$^{+4\mathrm{E}-4}_{-4\mathrm{E}-4}$ & 0.55$^{+1\mathrm{E}-4}_{-1\mathrm{E}-4}$ & 15.30$^{+4\mathrm{E}-3}_{-8\mathrm{E}-4}$ & 2.26$^{+0.01}_{-0.01}$ & -0.03$^{+1\mathrm{E}-4}_{-1\mathrm{E}-4}$ & 0.01$^{+1\mathrm{E}-4}_{-1\mathrm{E}-4}$ \\
    \hline
\end{tabular}
\tablefoot{A full description of the model is given in Sec.~\ref{sec:visibility_mod}. Column (1): ALMA Band. Column (2): Gaussian peak intensity. Column (3): Standard deviation of the Gaussian. Column (4): Inclination of the Gaussian. Column (5): Position angle of the Gaussian. Column (6): Envelope peak intensity. Column (7): Inner radius of the envelope. Column (8): Outer radius of the envelope. Column (9): Power law exponent. Column (10): Offset in Right Ascension. Column (11): Offset in Declination.}
\end{table*}

\section{Synthetic observations test}
\label{appendix:test_synthetic_obs}

The dust emissivity index profiles derived in this work (Fig.~\ref{fig:beta_profile}), which combine single-dish and interferometric measurements, exhibit a discontinuity between the values obtained at large and small spatial scales for L1448-C. To determine whether this offset could arise from the methodology rather than reflect a genuine physical effect, we tested our analysis procedure using synthetic observations from \citet{Valdivia2019}. This simulation depicts an isolated core collapse without cloud scale emission.

From the synthetic maps of the 1.3 and 3.2 mm observations, we extracted radial intensity profiles following the same procedure as for the single-dish data (see Sec. \ref{subsec:radial_dust_emissivity_profiles}). For the synthetic visibilities, we removed the compact component contribution, by measuring the flux 160 k$\lambda$ and subtracting it from the visibilities (lower than the 200 k$\lambda$ used by \citep{Galametz2019}). From these corrected visibilities, we computed the spectral index profile. 

\begin{figure}
  \resizebox{\hsize}{!}{\includegraphics{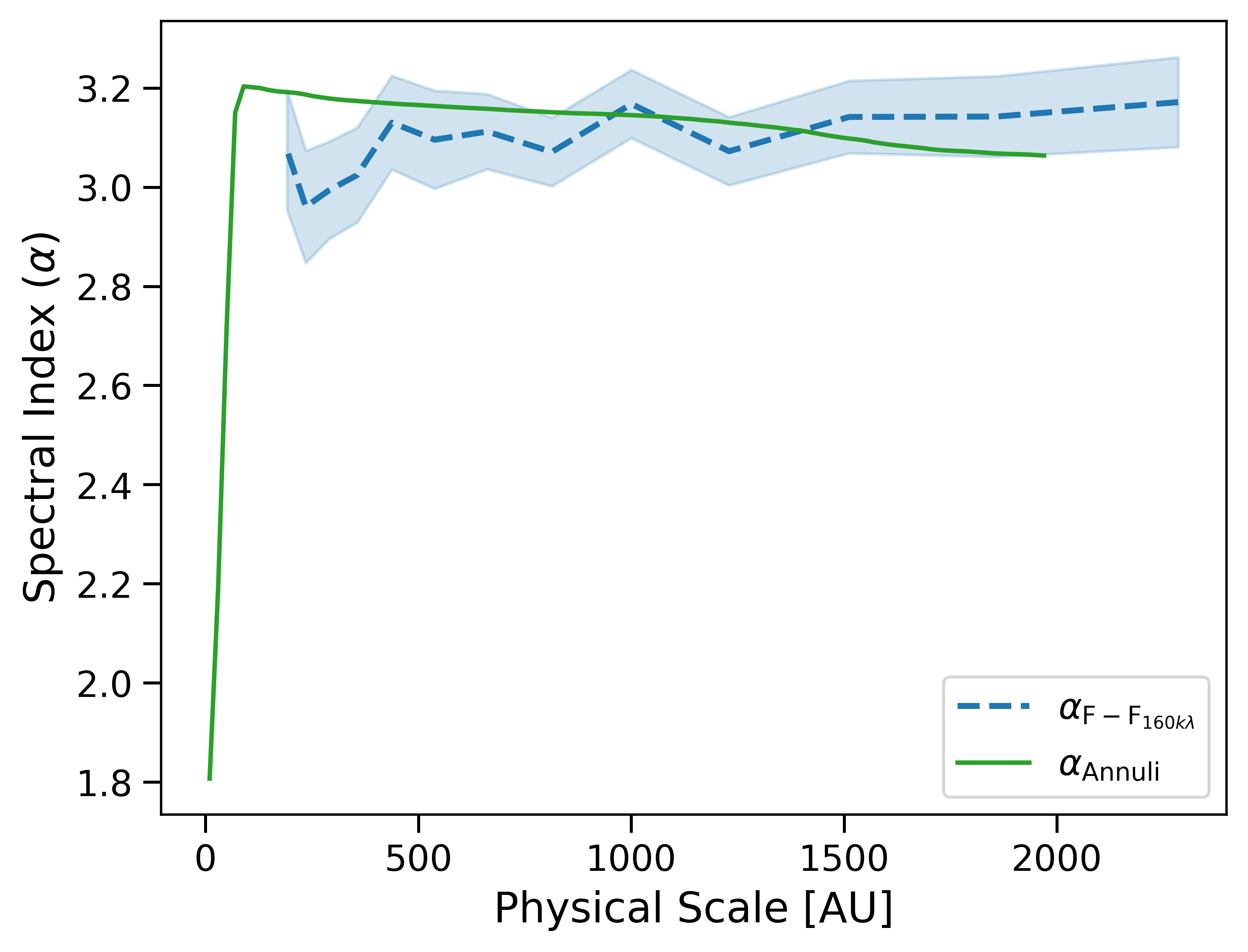}}
  \caption{Spectral index measured in the envelope model presented in \citep{Valdivia2019} from the image plane (green) and from the visibilities (dashed blue line).}
  \label{fig:test_synthetic_obs}
\end{figure}

The resulting spectral index profiles derived from both the synthetic single-dish and interferometric datasets are shown in Fig.~\ref{fig:test_synthetic_obs}. They are consistent with a flat distribution, showing no slope or offset between the two measurements. This demonstrates that our analysis procedure does not introduce gradients or discontinuities in $\alpha$ (and consequently in $\beta$), and that our method is not responsible for the offset we observe for L1448-C in Fig.~\ref{fig:beta_profile}.

\section{NIKA2 Transfer function}
\label{appendix:filtering}

The signal obtained by a single-dish antenna is filtered in a complex way. The amount of filtering depends on the atmospheric conditions and their stability at the time of the observations, on the brightness and geometry of the source, and on the observing parameters. Emission coming from diffuse extended regions can then be suppressed during the sky noise subtraction, as this emission is partly indistinguishable from the atmosphere and is therefore attenuated. In this appendix we quantify the presence and amplitude of this filtering in our NIKA2 data by characterizing the transfer function of the data reduction pipeline, and we propagate it to the quantity most sensitive to it in this work, the millimeter spectral index $\alpha_{\rm NIKA2}$.

To create noiseless sky models at both NIKA2 bands we started from the modified black body equation:

\begin{equation}
    I_{\nu} = \mu_{\rm H_{2}} m_{\rm H} N_{\rm H_{2}}\kappa_{\nu}B_{\nu}(T_{\rm d}),
    \label{eq:MBB_annex}
\end{equation}

where $\mu_{\rm H_{2}}$ is the mean molecular weight per $\rm H_{2}$ molecule, $m_{\rm H}$ is hydrogen mass, $N_{\rm H_{2}}$ the column density, $\kappa_{\nu}$ the dust opacity, and $B_{\nu}(T_{\rm d})$ the Planck function evaluated at $T_{\rm d}$. This is another form of Eq.~\ref{eq:MBB}. For the opacity we adopted $0.1~\mathrm{cm^{2}g^{-1}} (\nu/1~\mathrm{THz})^{\beta}$, with $\beta=1.6$. We used as an input the column density and temperature maps from \citet{Pezzuto2021}, which were then evaluated with Eq.~\ref{eq:MBB_annex} at $1.15~\rm mm$ and $2~\rm mm$. We note that these models are limited to 18\arcsec, inherited from the HGBS high resolution column density map, which is comparable to the resolution of NIKA2 at $2.00~\rm mm$.

The data reduction pipeline, PIIC, allows model maps to be injected into the data timeline, so that the same reconstruction process acts on the synthetic data. We injected the two synthetic maps and processed them with the same parameters used for the NIKA2 science maps. Writing the pipeline as a linear filter acting on the sky plus a term uncorrelated with the input,
 
\begin{equation}
    \tilde{I}_{\rm out}(\mathbf{k}) = T(k)\,\tilde{I}_{\rm mod}(\mathbf{k}),
    \label{eq:filter}
\end{equation}
 
the transfer function is estimated from the cross spectrum,

\begin{equation}
    T(k) = \frac{\left\langle \tilde{I}^{*}_{\rm mod}(\mathbf{k})\,
                 \tilde{I}_{\rm out}(\mathbf{k})\right\rangle}
                {\left\langle |\tilde{I}_{\rm mod}(\mathbf{k})|^{2}\right\rangle}.
    \label{eq:TF}
\end{equation}

Here $\tilde{I}_{\rm mod}$ and $\tilde{I}_{\rm out}$ are the Fourier transforms of the injected noiseless model and of the map reconstructed by PIIC, $\mathbf{k}$ is the spatial frequency, corresponding to an angular scale $\theta=1/k$. The asterisk denotes complex conjugation and $\left\langle\cdot\right\rangle$ an average over the Fourier modes in each annulus of constant $k$.

\begin{figure}
\resizebox{\hsize}{!}{\includegraphics{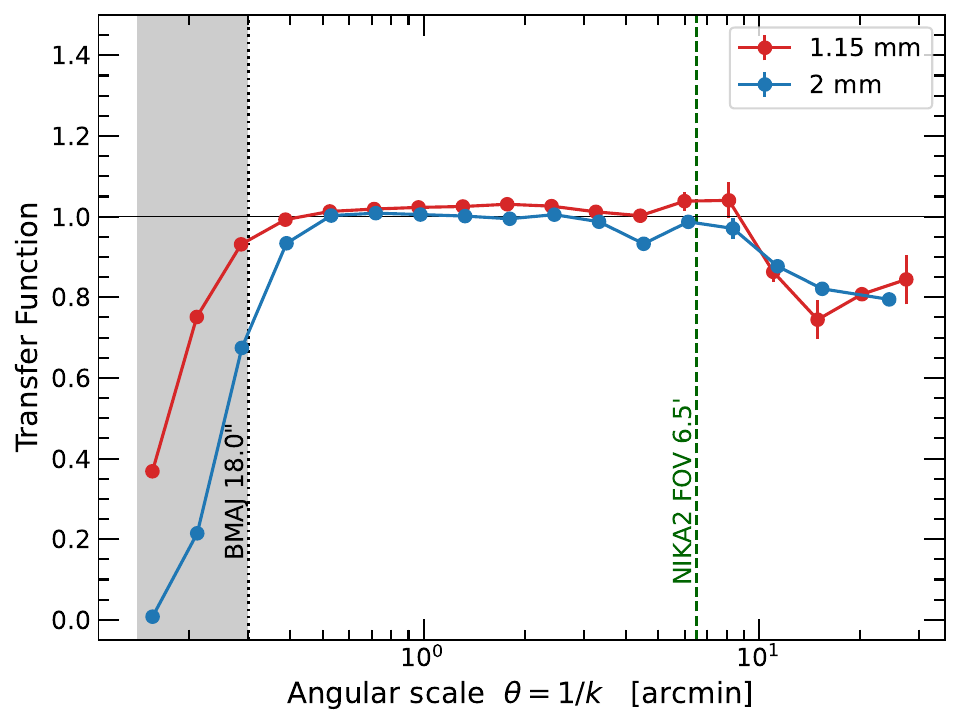}}
    \caption{Transfer function of the 1.15 mm and 2 mm maps, indicated by lines of different colors. Two angular scales are shown as vertical lines in both maps: the 18\arcsec scale of the input model is shown as a dotted black line, and the instrument 6.5\arcmin~FOV is shown as a green dotted line.}
  \label{fig:transfer_function}
\end{figure}

We show the resulting transfer functions in Fig.~\ref{fig:transfer_function}. The transfer function is consistent with unity at the scales of interest. Both bands remain within $5\%$ of $T=1$ for $0.4'\lesssim\theta\lesssim8'$, fall below $0.9$ only for $\theta\gtrsim 10$\arcmin, and reach $T\simeq0.8$ at the largest scale measured. At the $6.5$\arcmin~FOV scales, $T=1.04\pm0.02$ at $1.15~\rm mm$ and $0.99\pm0.01$ at $2~\rm mm$. We expect this value close to unity as the scan lengths are considerably longer than the array FOV. Below the $18$\arcsec~scales, the measured $T$ collapses in both bands. This is an effect of the resolution in the input model. This behavior was also observed in \cite{Zhou2026}, who attributed it to model resolution, the interaction of PIIC with the model, and potential errors with the creation of null maps.

As we have characterized the transfer function in both NIKA2 bands, we can estimate its effect on the derived $\alpha_{\rm NIKA2}$ as this quantity responds directly to the ratio between both bands, and can incorporate it to the global error budget.

\begin{equation}
    \Delta\alpha(k) = \frac{\ln\left[T_{1.15\,\rm mm}(k)/T_{2\,\rm mm}(k)\right]}
                           {\ln\left(\nu_{1.15\,\rm mm}/\nu_{2\,\rm mm}\right)},
    \label{eq:dalpha}
\end{equation}

Over the range of scales probed by our observations, bounded by the 36.3\arcsec common beam of the SED and by the $\sim8$\arcmin extent of the emission, $\Delta\alpha$ ranges from $0.02$ to $0.12$, with a power-weighted mean of $0.06$. This is a bias and not an uncertainty, however, since the amplitude depends on angular scales that varies along the map, we opt for incorporating it as an additional systematic uncertainty of $0.06$ on $\alpha_{\rm NIKA2}$, added in quadrature to the global error budget.

We further tested the effect of spatial filtering on our analysis by applying the transfer function onto the Herschel data and repeating the SED fitting and computation of $\beta_{\rm NIKA2}$. The median difference in $\beta_{\rm NIKA2}$ with respect to the unfiltered fit is 0.04, which is within the systematic uncertainty quoted above. The overall conclusion is that spatial filtering is not a limiting effect on the results presented in this work.

\end{appendix}
\end{document}